\documentclass[10pt,journal,compsoc]{IEEEtran}

\usepackage{main_header}

\AtBeginDocument{%
  \providecommand\BibTeX{{%
    \normalfont B\kern-0.5em{\scshape i\kern-0.25em b}\kern-0.8em\TeX}}}

\begin{document}
\title{Collision Prediction from UWB Range Measurements }

\author{Alemayehu Solomon Abrar, Anh Luong, Gregory Spencer, Nathan Genstein, Neal Patwari and\\ Mark Minor

\thanks{ Alemayehu Solomon Abrar was with the Preston M. Green Department of Electrical \& Systems Engineering at Washington University in St. Louis}

\thanks{ Anh Luong was with Carnegie Mellon University}

\thanks{ Gregory Spencer was with the University of Utah}

\thanks{ Nathan Genstein was with the Preston M. Green Department of Electrical \& Systems Engineering at Washington University in St. Louis}

\thanks{ Neal Patwari is with McKelvey School of Engineering at Washington University in St. Louis}

\thanks{ Mark Minor was with the University of Utah}

}

\IEEEtitleabstractindextext{%
\begin{abstract}
The ability to predict, and thus react to, oncoming collisions among a set of mobile agents is a fundamental requirement for safe autonomous movement, both human and robotic.  This paper addresses systems that use range measurements between mobile agents for the purpose of collision prediction, which involves prediction of the agents' future paths to know if they will collide at any time. One straightforward system would use known-location static anchors to estimate agent coordinates over time, and use the track to predict collision.  Fundamentally, no fixed coordinate system is required for collision prediction, so using only the pairwise range between two agents can be used to predict collision.  We present lower bound analysis which shows the limitations of this pairwise method.  As an alternative anchor-free method, we propose the friend-based autonomous collision prediction and tracking (FACT) method that uses all measured ranges between nearby (unknown location mobile) agents, in a distributed algorithm, to estimate their relative locations and velocities and predict future collisions between agents.  Using analysis and simulation, we show the potential for FACT to achieve equal or better collision detection performance compared to other methods, while avoiding the need for anchors.  We then build a network of $N$ ultra wideband (UWB) devices and an efficient multi-node protocol which allows all ${\cal O}(N^2)$ pairwise ranges to be measured in $N$ slots.  We run experiments with up to six independent robot agents moving and colliding in a 2D plane and up to four anchor nodes to compare the performance of the collision prediction methods.  We show that the FACT method can perform better than either other method but without the need for a fixed infrastructure of anchor nodes.


\end{abstract}


\begin{IEEEkeywords}
Collision Prediction, Autonomous, Ultra wideband,
\end{IEEEkeywords}
}

\maketitle

\section{Introduction}
\label{sec:intro}

\IEEEPARstart{U}{nexpected} collisions that occur within a set of mobile agents, be it mid-air collisions or traffic collisions, \replace{are usually}{can be} devastating. In contact sports like American football, head collisions are the main causes of millions of traumatic brain injuries such as concussions every year \cite{daneshvar2011epidemiology}. We investigate smart sensor-enabled devices which predict an impending collision between agents and take action to avoid or lessen the damage from collision, for example, smart helmets that alert a player before a collision from behind, or autonomous drones that avoid colliding.


Such systems  rely heavily on reliable sensing.   Sensing modalities  for this purpose including global navigation satellite system (GNSS) tracking \cite{remote_2019}, inertial measurements, vision sensors, and ultrasonic and millimeter wave radar systems. 
%
Existing approaches have been used to enable key safety features in particular applications, but do have shortcomings.
GNSS-based systems require significant battery power, can be very inaccurate in urban areas, and unavailable indoors. Vision-based systems have shortcomings when the impending collision is between small or fast objects, and in poor light conditions. Radar systems must use high transmit power to contend with $d^{-4}$ losses. 

In this paper, we investigate autonomous objects which predict (and react to) impending collisions among themselves using pairwise range measurements.  We note that UWB transceivers provide low cost and energy-efficient range measurements with fine resolution, on the order of a few cm of standard deviation, due to their use of very narrow pulses. Range-based collision sensing is complementary to existing approaches because it is reliable in both indoor and outdoor environments, for small or large objects, and in the dark or in the light.  Two-way ranging has $d^{-2}$ (as opposed to $d^{-4}$) losses, enabling fast moving objects to predict a collision even from a long distance. Such systems require each object to have a transceiver, but as opposed to proposed FCC GNSS requirements for drones \cite{remote_2019}, pairwise ranging systems can autonomously avoid collision and can be lower in power consumption.

\paragraph{Localization vs.\ Collision Prediction}
 Unlike localization, collision prediction does not require knowledge of absolute coordinates. Whether or not a collision will occur depends on only the \emph{relative} kinematics of objects, including relative position and relative velocity. Thus, it is not necessary to have infrastructure nodes or a global coordinate reference. Compared to localization, however, collision prediction is significantly more challenging because it must predict future positions for all time from now until a future time $\Delta t$ from now.  To be clear about this important point: predicting position at only the time $\Delta t$ from now is not sufficient --- an impending collision should be predicted if two trajectories will intersect \emph{for any time between now and $\Delta t$ later}. As such, collision prediction depends critically on accurate estimation of relative velocity (and relative acceleration), and it always involves two or more nodes.  In comparison, localization does not require kinematics or multiple nodes.

Despite these differences, a na\"ive approach for collision prediction could simply first estimate absolute coordinates and track for each individual agent, predict each agent's future trajectory, and determine if these coordinate trajectories intersect.
A typical approach in localization with UWB range measurements involves multilateration of a tag node with respect to static reference nodes called anchors \cite{weiser2016characterizing}. However, anchor-based localization entails the need for deployed infrastructure of several anchors, each with location determined with some other method.

\paragraph{Relative Localization and Tracking}
Based on the fact that collision prediction requires only relative kinematics, our proposed method is more closely related to past methods in relative localization. 
A classical relative approach is to use multi-dimensional scaling (MDS) to generate a relative map. However, such relative position estimates cannot be directly used to determine relative velocity since each successive position estimate has an arbitrary translation, rotation, and flip. Rajan et al.\ proposed to determine relative velocity with a modified MDS method in which relative velocity is estimated from the second order derivative of squared distance measurements \cite{rajan2019relative}. Yet, this method results in inaccurate velocity estimates in the presence of noisy range measurements. 

\paragraph{Questions on Performance} 
For some applications, the requirement for anchors will be infeasible, and in others, pairwise range measurements will be infeasible.  Yet the question remains: Does anchor-based localization provide better collision prediction performance than a system using only relative range measurements between nodes?  This question is largely unanswered in the literature.  This paper provides analysis and data that address this question.

A second key question addressed in this paper is:  For methods using only relative range measurements, does increasing the $N$ used by a collision predictor improve its performance?  Minimally, $N=2$ nodes $i$ and $j$ can measure their relative distance over time, and use the measurements to predict an impending collision between $i$ and $j$, which we call the pairwise method.  We propose to additionally use measurements between pairs of nodes in the  set including $i$, $j$, and other ``friend'' nodes.  We are unaware of any published relative collision prediction method using $N>2$. Although no node has absolute coordinate information, we show that the additional relative range measurements can improve the prediction of collision between $i$ and $j$, via analysis and experiment.

\paragraph{Contribution 1: FACT Algorithm}
In this paper, we propose and evaluate a novel collision prediction method that uses only relative range measurements and distributed computation for mobile nodes, which we call the Friend-based Autonomous Collision prediction and Tracking (FACT) method.  The FACT method assumes each node is equipped with a UWB transceiver for ranging, that $N$ mobile neighboring nodes cooperate to compute all pairwise range measurements, and use them all to predict the impending collision of each pair of nodes.  We evaluate and compare the method with pairwise and anchor-based approaches via theoretical bound analysis, simulation, and experiments involving robotic vehicles.

\paragraph{Contribution 2: $\mathcal{O}(N)$ Multi-node Ranging Protocol \& Implementation}
We found, however, that existing UWB protocols did not allow us to measure $N\choose 2$ pairwise range measurements at a high enough rate when $N$ is large.  In existing protocols, many UWB packets are used solely to estimate clock frequency offset (which has more significant effects in multi-node protocols).  Existing protocols use $\mathcal{O}(N^2)$ UWB packets to separately measure the pairwise ranges \cite{kempke2015polypoint}.  Note that the UWB time stamp data increases as $\mathcal{O}(N^2)$ when sharing $N$ time stamps for each of $N$ nodes.  In combination, the range measurement period is long and rises as $\mathcal{O}(N^2)$.  

Our key insight is that using a secondary transceiver, capable of handling both frequency synchronization and data sharing, allows us to implement a dramatically faster multi-node ranging protocol.  As narrowband (as compared to UWB) transceivers can use much higher transmit powers, they can communicate data with much greater bandwidth efficiency.  Further, they can be used to synchronize local oscillators among all nodes.  In our implementation, we develop and test our multi-node ranging protocol which, using a narrowband transceiver for synchronization and time stamp data communication, dramatically shortens the measurement period.  In one round of our multi-node ranging protocol, each node exchanges a single short UWB packet, which enables each node to calculate $N \choose 2$ ranges.  Thus the measurement period for calculating $N \choose 2$ ranges reduces to $\mathcal{O}(N)$ from the $\mathcal{O}(N^2)$ of prior protocols.




%
\section{Problem Statement}
\label{sec:problem}

The objective of collision prediction is to detect a future intersection between the trajectories of a pair of mobile agents. We assume that, for a short period of time, node velocities can be well approximated as constant. 

\begin{figure}[htpb]
  \begin{center}
     \includegraphics[width=0.7\columnwidth]{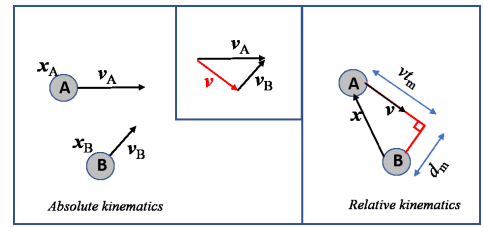}
  \caption{Agents with linear relative motion}
  \label{fig:assumption}
\end{center}
\end{figure}

Consider, for example, a simple scenario where two nodes A and B move in uniform rectilinear motion. We assume that node A and node B, initially located at $\textbf{x}_A$ and $\textbf{x}_B$ move at constant velocities  $\textbf{v}_A$ and $\textbf{v}_B$ respectively as shown in Figure \ref{fig:assumption}.  For the given setup problem, we are interested in the relative position $\textbf{x}_{BA}$ and relative velocity $\textbf{v}_{BA}$. For convenience, we drop subscripts for all subsequent relative vectors with respect to node A.  Their relative initial distance and speed are denoted by $x$ and $v$ respectively. 

We determine the possibility of collision between a pair of non-stationary agents based on three scalar \emph{collision prediction (CP)} parameters, namely: 
	\begin{itemize}
	    \item $v$: relative speed
	    \item $d_m$: the minimum passing distance between two nodes.
	    \item $t_m$: the time at which the relative distance reaches its minimum. 
	\end{itemize}

For a given initial relative position and relative velocity, the collision prediction parameters can be written as:
\begin{equation}
    \begin{aligned}
        v &= \lVert \textbf{v}\lVert\\
        t_m &= -\dfrac{\textbf{x}\cdot\textbf{v}}{v ^2} \\
        d_m &= \dfrac{1}{v}\left({\lVert {\textbf{x}} \lVert^2\lVert {\textbf{v}} \lVert^2 - \left({\textbf{x}\cdot\textbf{v}} \right)^2}\right)^{\frac{1}{2}}
    \end{aligned}
    \label{eqn:param}
\end{equation}

In practical settings, a node has a non-zero volume, and can be modelled by a spherical volume with some radius $r$. In this case, collisions could occur even prior to reaching the minimum passing distance. Assuming all nodes with the same radius $r$ moving in a 2-D plane, then for positive $t_m$ and $d_m<2r$, the time of collision $t_c$ is calculated as:
\begin{equation}
    t_c=t_m-\dfrac{1}{v}\sqrt{4r^2-d_m^2}
\end{equation}

The values of CP parameters estimated for a pair of mobile agents can be used to determine if there is a future collision. For example, a system may declare an inbound collision when two nodes are coming close to less than a meter ($d_m<1m$) within half a second ($t_m<0.5s$)
Furthermore, the accuracy of a collision prediction method can be evaluated in terms of its performance in accurately estimating of these parameters.

In this article, we present methods to estimate CP parameters solely from distance measurements to predict collisions. 

\section{Estimation and Detection}
\label{sec:est_bound}
In this section, we present methods to estimate collision prediction parameters and  to detect oncoming collisions using range measurements. We develop statistical bounds based on Cram\'er-Rao lower bound (CRLB) analysis to evaluate the methods.

Range measurements between a set of mobile agents can be used in distributed or an infrastructure-based manner to estimate CP parameters. Distributed collision prediction is performed using range measurements from either a single or multiple pairs simultaneously.
We refer to methods that use range measurements from a single pair as pairwise methods and those employing multiple pairs for collision prediction of every pair
as Friend based methods. On the other hand, for ranging involving nodes taking measurements with respect to fixed reference nodes, we employ anchor-based collision prediction.   

\subsection{ Cram\'er-Rao Lower Bound Analysis}
We apply CRLB analysis in order to determine theoretically maximum accuracy for methods used in collision prediction.  The CRLB is the most common variance bound due to its simplicity \cite{kay}.  It provides the lowest possible estimation variance achieved by any unbiased estimator. To the best of our knowledge, we present the first bound analysis in relation to collision prediction from range measurements.
We compute the bound on estimation variance for a CP parameter vector given by $\bm{\theta} \coloneqq \left[d_m, t_m, v\right]^T$ under multiple range measurement models.

\subsubsection{Bounds for Pairwise Ranging}

For pairwise methods, the CP parameters are estimated using range measurements from a single pair of nodes. We consider a noisy distance $\delta(t_n)$ between the nodes measured at time $t_n$, which is given by

\begin{equation}
    \begin{aligned}
        \delta(t_n) &=  \lVert\textbf{x} + \textbf{v}t_n\lVert + w(t_n)\\ 
       &= \left( {d_m^2 +v^2\left(t_m-t_n\right)^2} \right)^{\frac{1}{2}}+ w(t_n)
    \end{aligned}
    \label{eqn:pairwise}
\end{equation}
where $w$ represents zero-mean, additive white Gaussian noise, $w \sim {\cal N} \left(0, \sigma^2 \right)$
For a set of N i.i.d.\ distance measurements $\bm{\delta} = \{\delta(t_1),  \cdots, \delta(t_N) \}$, the log-likelihood becomes
\begin{equation}
    \begin{aligned}
        {\cal L}(\bm{\theta}|\bm{\delta}) 
        &=-\frac{1}{2}\log\left((2\pi)^N\sigma^2\right)\\
        \qquad &-\dfrac{1}{\sigma^2}\sum_{n=1}^{N}{\left(\delta(t_n)-\sqrt{d_m^2+v^2\left(t_m-t_n\right)^{2}}\right)^2}.
    \end{aligned}
\end{equation}
Defining ${\cal I}(\bm{\theta}) = - {\rm E}\left[ \partial^2 {\cal L}(\bm{\theta}|\bm{\delta}) / \partial \bm{\theta}^2 \right]$, we have \begin{equation}
       {\cal I}(\bm{\theta}) = \dfrac{1}{\sigma^2}\sum_{n=1}^{N}
        \dfrac{F_n}{d_m^2 +v^2\left(t_m-t_n\right)^2}
    \label{eqn:pairfim}
\end{equation}
where 
\begin{equation}
       F_n  =
       \begin{bmatrix}
           v^4(t_m-t_n)^2   & v^2d_m(t_m-t_n)   & v^3(t_m-t_n)^3 
            \\
           v^2d_m(t_m-t_n)  & d_m^2           & vd_m(t_m-t_n)^2
            \\
            v^3(t_m-t_n)^3  & vd_m(t_m-t_n)^2 & v^2(t_m-t_n)^4
    
        \end{bmatrix}
    \nonumber
\end{equation}
The theoretical bounds on variance for $t_m$,  $d_m$ and $v$ estimates are computed as:
 \begin{equation}
    \begin{aligned}
       \mbox{var}(\hat{t}_m)&\geq \left\{{\cal I}(\bm{\theta})^{-1}\right\}_{11}  \\ 
       \mbox{var}(\hat{d}_m)&\geq \left\{{\cal I}(\bm{\theta})^{-1}\right\}_{22} \\
       \mbox{var}(\hat{v})&\geq \left\{{\cal I}(\bm{\theta})^{-1}\right\}_{33}
    \end{aligned}
    \label{eqn:bound}
\end{equation}

\subsubsection{Bounds for Anchor-Based Ranging}
\label{sec:anchorcb}
In anchor-based ranging, every node takes measurements with respect to nodes with fixed known location. we assume that there are $K$ anchor nodes with the $k^\text{th}$ anchor node located at $\textbf{x}_k$. For a pair having $i^{th}$ and $j^{th}$ mobile nodes, the distances with respect anchor $k$ are denoted by $x_{ik}$ and $x_{jk}$ respectively. The measured distance vector at anchor $k$ is given by:
 \begin{equation}
    \begin{aligned}
        \bm{\delta}_k(t_n) = 
        \begin{bmatrix}
             \delta_{ik}(t_n) \\
             \delta_{jk}(t_n)
        \end{bmatrix}
        &= \begin{bmatrix}
             x_{ik}(t_n) \\
             x_{jk}(t_n)
        \end{bmatrix} + \textbf{w}_k(t_n)
        \\
        &= \bm{\bar{\delta}}_k(t_n) + \textbf{w}_k(t_n)
    \end{aligned}
\end{equation}
 Where the additive noise is assumed to be Gaussian (i.e, $\textbf{w}_k \sim {\cal N}(\textbf{0}, \sigma^2\textbf{I})$
 
  

Since there is no direct relationship between the parameters in $\bm{\theta}$ and the range measurements $\bm{\delta}_k$, We define initial positions and velocities of a pair of nodes as the initial parameters. Hence, the initial parameter vector is given by $\bm{\alpha} \coloneqq \left[\textbf{x}_i, \textbf{x}_j, \textbf{v}_i, \textbf{v}_j\right]^T$. The log-likelihood distribution for measurement vector $\bm{\delta}_k$ at the $k^\text{th}$ anchor becomes
 \begin{equation}
     \log{p(\bm{\delta}_k(t_n)=\textbf{d}_k|\bm{\alpha})} = \ -\left(\left(\dfrac{\lVert\textbf{d}_k-{\bm{\bar{\delta}}_k(t_n)}\Vert^2}{2\sigma^2}\right)+\log{\left(2\pi\sigma^2\right)}\right)
     \nonumber
 \end{equation}
Assuming i.i.d.\ distance measurements, the log likelihood for all measurements from $K$ anchors for $N$ samples is calculated as
\begin{equation}
    {\cal L}(\bm{\alpha}|\bm{\delta}) = \sum_{k=1}^{K} \sum_{n=1}^{N} \log  p(\bm{\delta}_k(t_n)=\textbf{d}_k|\bm{\alpha}) 
\end{equation}

If we define $\textbf{R}_{ik}(t_n) \coloneqq \dfrac{ {\textbf{x}_{ik}(t_n)}{\textbf{x}_{ik}(t_n)}^T}{\lVert{\textbf{x}_{ik}(t_n)}\lVert^2}$, then ${\cal I}(\bm{\alpha})$ becomes
           
          
           
\begin{equation}
       {\cal I}(\bm{\alpha}) =  \dfrac{1}{\sigma^2}\sum_{k=1}^{K}\sum_{n=1}^{N} {\bm{\Gamma}}_k(t_n).
    \nonumber
\end{equation}
where
\begin{equation}
       {\bm{\Gamma}}_k(t_n) =
       \begin{bmatrix}
           \textbf{R}_{ik}(t_n) &\textbf{0} &t_n\textbf{R}_{ik}(t_n) &\textbf{0}\\
           
           \textbf{0}  & \textbf{R}_{jk}(t_n) &\textbf{0} &t_n\textbf{R}_{jk}(t_n)\\
          
           t_n\textbf{R}_{ik}(t_n) &\textbf{0} &t_n^2\textbf{R}_{ik}(t_n) &\textbf{0}\\
           
           \textbf{0} &t_n\textbf{R}_{jk}(t_n) &\textbf{0} &t_n^2\textbf{R}_{jk}(t_n)
        \end{bmatrix}.
    \nonumber
\end{equation}

The desired parameter vector $\bm {\theta} = [d_m, t_m, v]^T$ is a function of the initial parameter vector $\textbf{g}(\bm{\alpha})$ as given by (\ref{eqn:param}). Then, we perform FIM transformation \cite{kay} of ${\cal I}(\bm{\alpha}) $ to ${\cal I}(\bm{\theta})$: 
\begin{equation}
    \left({\cal I}(\bm{\theta})\right)^{-1} = \left({\frac{\partial}{\partial\bm{\alpha}}\textbf{g}(\bm{\alpha})}\right){\cal I}(\bm{\alpha})^{-1}\left({\frac{\partial}{\partial\bm{\alpha}}\textbf{g}(\bm{\alpha})}\right)
\end{equation}
 Which leads to the theoretical bounds on estimation variance for $t_m$,  $d_m$ and $v$ as given in (\ref{eqn:bound}).

\subsubsection{Bounds for Friend-based Ranging}
\label{sec:friendcrb}
When there are no anchors, measurements between a given pair and those with respect to other mobile nodes can be used together to improve estimation of CP parameters. In this setup, the measurement vector corresponding to the pair having the $i^{th}$ and $j^{th}$ nodes has two parts: the first part involves range measurements of the pair with respect to each friend node. In the analysis, we ignore range measurements between friend nodes other than $i^{th}$ and $j^{th}$ node as the the CP parameters are not related with these measurements and the FIM resulted from these measurements is zero.
For the $i^{th}$ and $j^{th}$ mobile nodes, the distances with respect $k^{th}$  friend are given by $x_{ik}$ and $x_{jk}$ respectively. Then, the distance vector corresponding the $k^{th}$ friend node at time $t_n$ is modeled as

\begin{equation}
    \begin{aligned}
        \bm{\delta}_k(t_n) = 
        \begin{bmatrix}
             \delta_{ik}(t_n) \\
             \delta_{jk}(t_n)
        \end{bmatrix}
        &= \begin{bmatrix}
             x_{ik}(t_n) \\
             x_{jk}(t_n)
        \end{bmatrix} + \textbf{w}_k(t_n)
        \\
        &= \bm{\bar{\delta}}_k(t_n) + \textbf{w}_k(t_n)
    \end{aligned}
\end{equation}
 Where the additive noise is assumed to be i.i.d.\ Gaussian (i.e., $\textbf{w}_k \sim {\cal N}(\textbf{0}, \sigma^2\textbf{I})$
 The second part of the measurement include the range measurements $\delta(t_n)$ between the pair of interest (i.e. the $i^{th}$ and $j^{th}$ nodes), and this is given in (\ref{eqn:pairwise}). 

We assume that all measurements are i.i.d. Hence the Fisher information matrix of CP parameters for friend-based ranging is given as a sum of two matrices corresponding to the two parts of the measurement. The first matrix is computed in the same as Section \ref{sec:anchorcb} except that the anchor positions are replaced by the unknown friend locations. The second matrix is based on pairwise measurements and given in (\ref{eqn:pairfim}).
\begin{equation}
    {\cal I}(\bm{\theta}) = {\cal I}_1(\bm{\theta}) + {\cal I}_2(\bm{\theta}) 
\end{equation}

The resulting FIM leads to bounds on estimation variance of the CP parameters using (\ref{eqn:param}).

\subsubsection{Numerical results} \label{sec:crb_numerical}

Next, we compare the three different approaches: pairwise, anchor-based, and friend-based, in terms of estimation variance bounds on the CP parameters.  We use simulation to provide a random set of node geometries. We first consider nodes initially positioned randomly within a circle of radius 50~m from the origin. Further, the nodes move at random constant velocity with speed in [0,10]~m/s. The anchor nodes are randomly positioned on a circle of radius 50~m. We run 100 trials and compute the bounds for the CP parameters under different measurement models.

\begin{figure}
  \begin{center}
     \includegraphics[width=0.6\columnwidth]{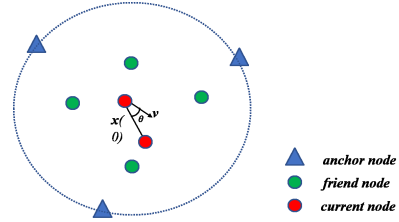}
  \caption{Simulation Setup}
  \label{fig:sim}
\end{center}
\end{figure}

\begin{figure}[h]
    \centering
    \begin{minipage}{0.4\textwidth}
     \includegraphics[width=1\textwidth]{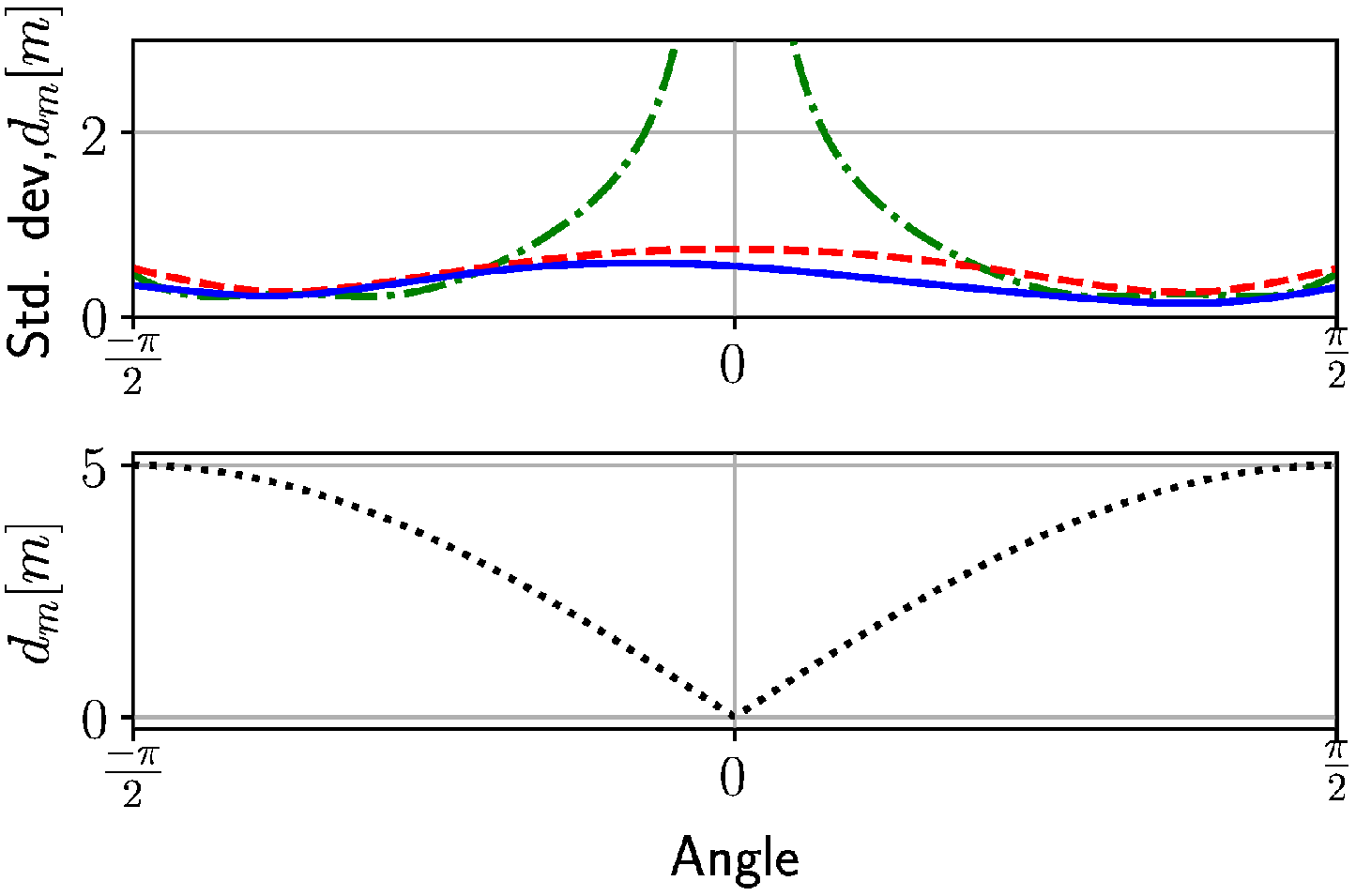}
  \end{minipage}
    \begin{minipage}{0.4\textwidth}
    \centering
     \includegraphics[width=1\textwidth]{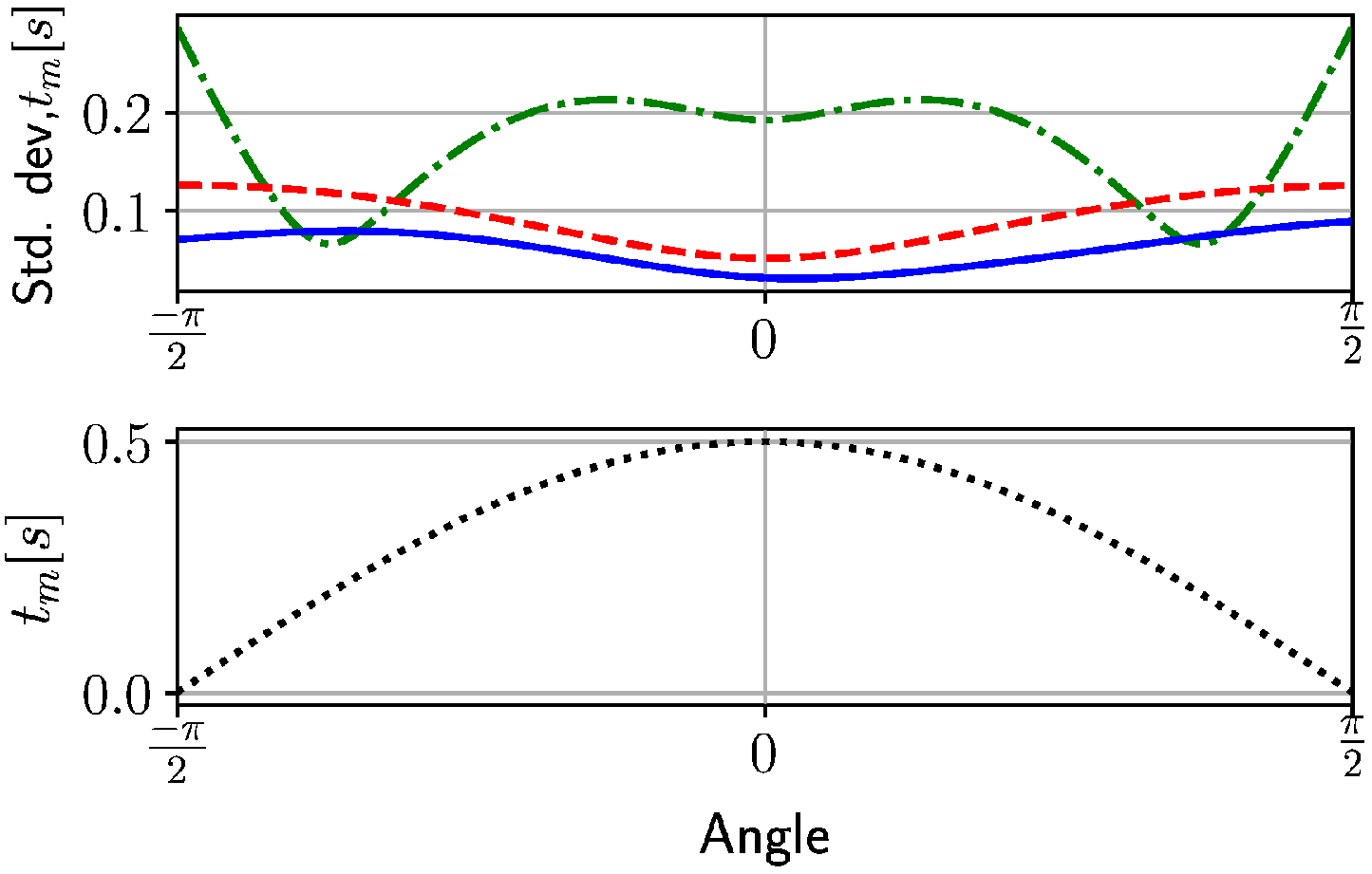}
  \end{minipage}
  \begin{minipage}{.4\textwidth}
  \centering
     \includegraphics[width=1\textwidth]{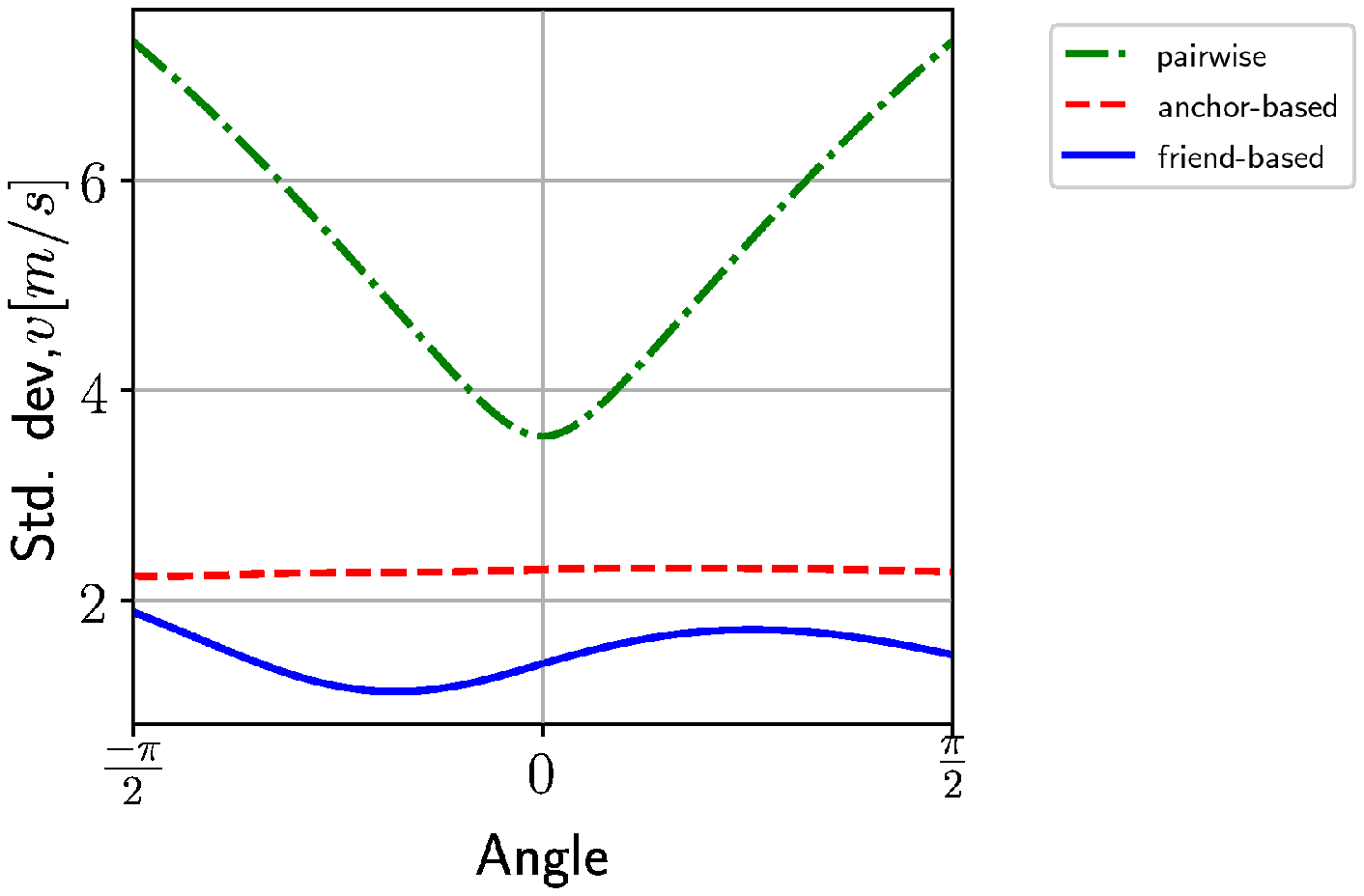}
  \end{minipage}
  \caption{Bound on standard deviation of estimation  for CP parameters}
  \label{fig:crb}
  \end{figure}

The bounds on standard deviation for estimated CP parameters are shown in Figure~\ref{fig:crb} considering the same number of anchor and friend nodes.
For a given initial position and speed of the nodes, we compute CRLB for varying angles of incidence.
The ranging noise is assumed to be white Gaussian with standard deviation of 0.1~m. We use 4 anchor and 4 friend nodes in the simulation.
We note from Figure \ref{fig:crb}, the pairwise ranging generally results in higher estimation standard deviation due to relatively few number of measurements. Particularly, for pairwise measurement, the standard deviation bound on $\hat{d}_m$  becomes large as the relative angle of incidence .  
On the other hand, friend-based ranging provides the lowest estimation variance of all the CP parameters. Anchor based ranging results in the second lowest estimation variance.

Compared to anchor based methods which use 2K measurements for each pair of nodes, friend-based methods are based on ${N\choose 2}$ measurements for $N$ mobile and $K$ anchor nodes. Friend-based ranging can be shown to improve estimation by lowering the  theoretical bound on the variance of $t_m$, $d_m$, and $v$ estimates despite the fact that friend node positions are unknown. Friend-based methods could thus improve collision prediction performance in anchor infrastructure-free systems. Furthermore, it can be shown that estimation variance increases with ranging noise and decreases with the number of friends and anchors in friend-based and anchor-based range respectively.

\begin{figure}[h]
    \centering
    \begin{minipage}{0.4\textwidth}
     \includegraphics[width=1\textwidth]{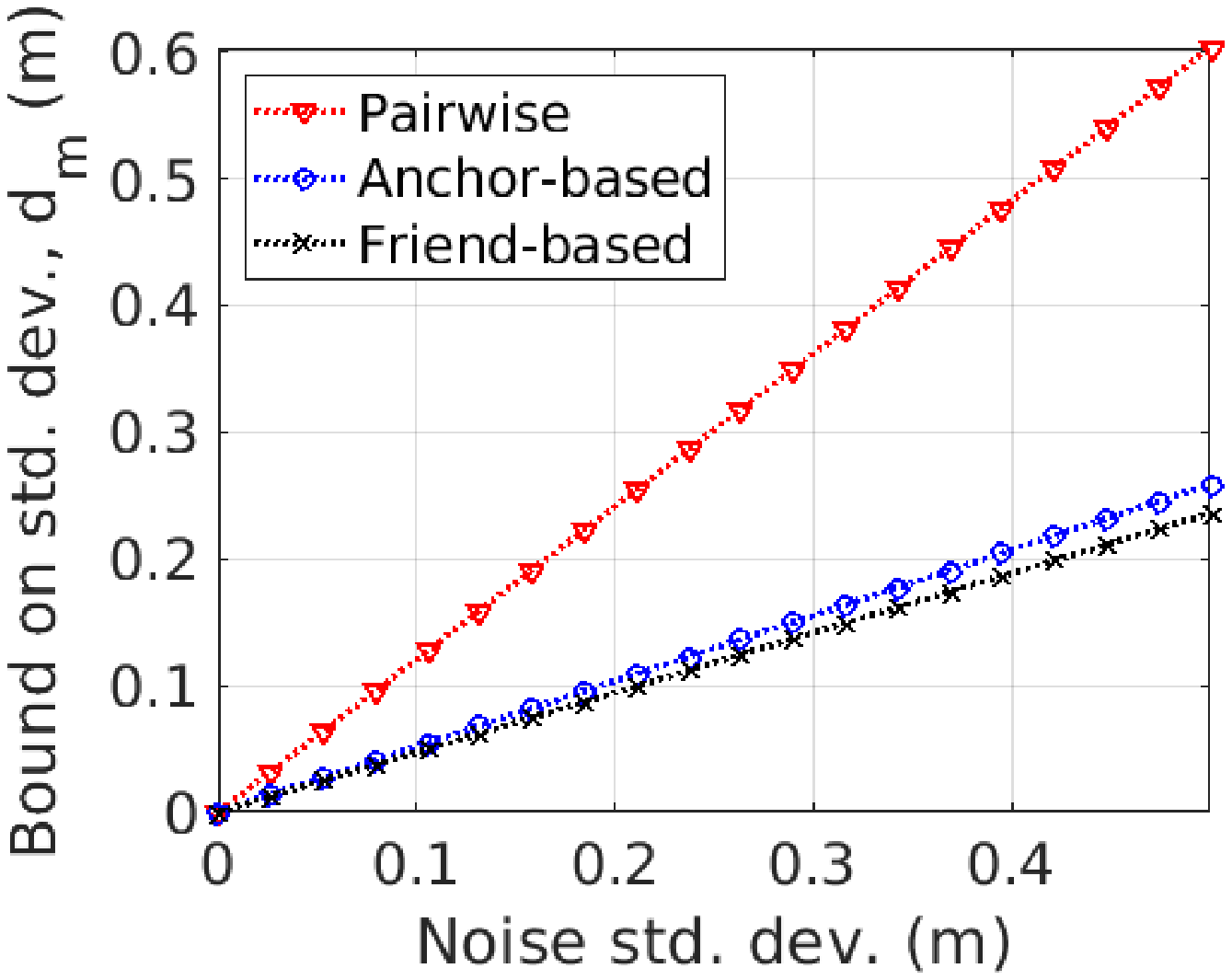}
  \end{minipage}
    \begin{minipage}{0.4\textwidth}
    \centering
     \includegraphics[width=1\textwidth]{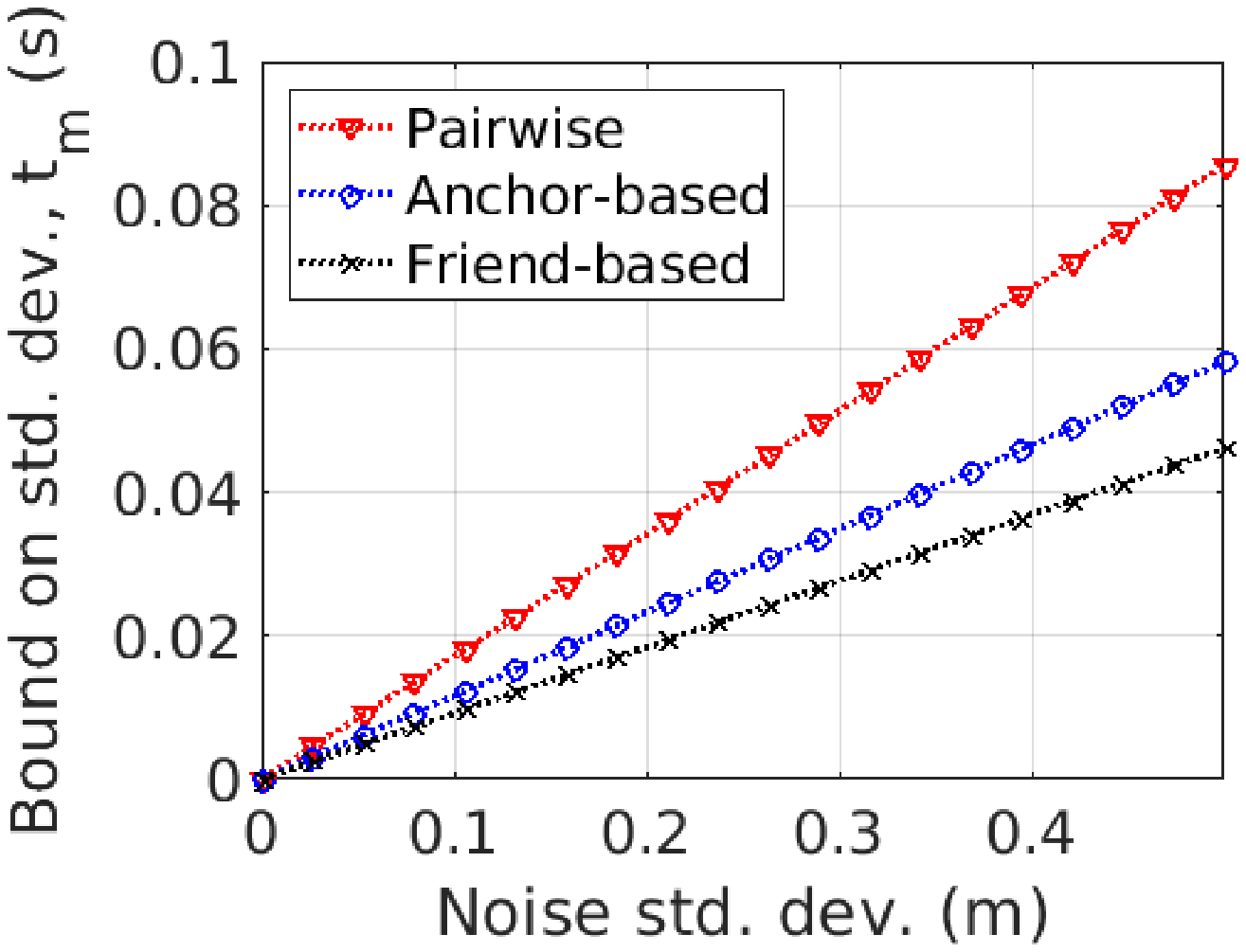}
  \end{minipage}
  \begin{minipage}{.4\textwidth}
  \centering
     \includegraphics[width=1\textwidth]{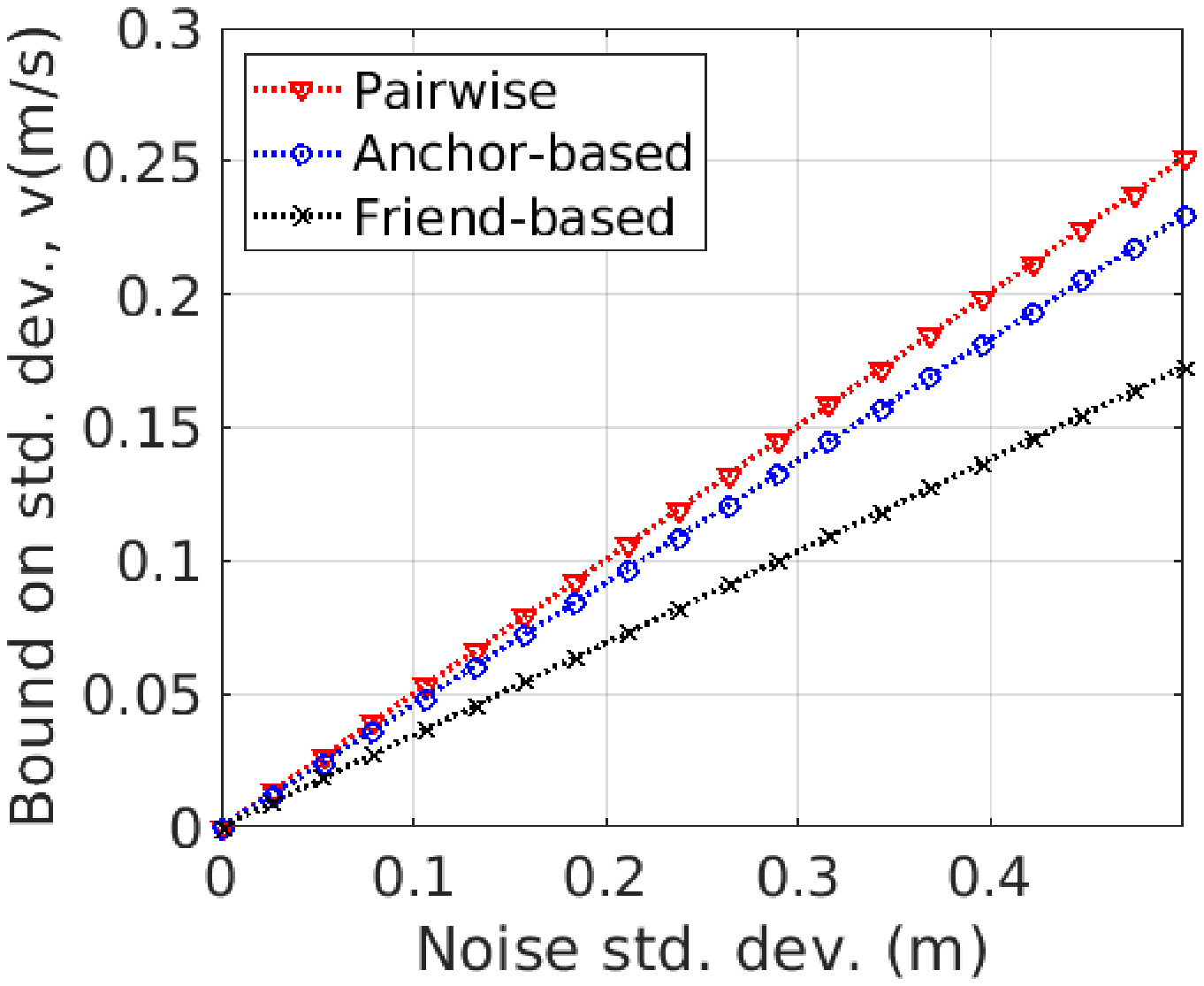}
  \end{minipage}
  \caption{Bound on standard deviation of estimation  as a function of ranging noise}
  \label{fig:crb_noise}
  \end{figure}

We also compare the estimation bounds for randomly generated geometries. In Figure \ref{fig:crb_noise}, we show the bound on estimation standard deviation for collision prediction parameters as a function of standard deviation of ranging noise. Every point in the plots is derived from 150 different geometries in 100 trials. The results show that for the estimation variances of the CP parameters $d_m$, $t_m$ and $v$ all increase with ranging noise, and friend-based approaches provide the lowest estimation bound in all cases. For example, for a ranging noise with standard deviation of 0.2m, the bound on $d_m$ for friend-based method is only 0.09m while anchor-based and pairwise methods provide bounds of 0.11m and 0.25m respectively.

\subsection{Collision Prediction Algorithms}
We develop three different methods to determine the possibility of collision between a pair of mobile nodes using only range measurements. The objective of the algorithms is to estimate CP parameters.

\subsubsection{Pairwise collision prediction}
In pairwise collision prediction, every pair uses range measurements from a single pair. Assuming nodes with linear motion, the squared distance between a pair $d^2$ is a quadratic function of time and can be expressed in terms of CP parameters as
\begin{equation}
    \begin{aligned}
        d^2(t) &= d_m^2 +v^2\left(t_m-t\right)^2\\
            &= a_0 + a_1t + a_2t^2
    \end{aligned}
\end{equation}

We apply quadratic regression on a set of squared pairwise range measurements to determine CP parameters as follows:
\begin{equation}
    \begin{aligned}
    t_m  &= -\dfrac{a_1}{2a_2}\\
    d_m  &= \sqrt{a_0-\frac{a_1^2}{4a_2}}\\
     v   &= \sqrt{a_2}
    \end{aligned}
\end{equation}

\subsubsection{Anchor-based Collision Prediction}
Anchor-based collision prediction involves two steps. First, we use time-difference-of-arrival (TDOA) measurements to find the absolute positions of each node. Second, we apply linear regression on estimated position coordinates to determine velocity.
The CP parameters are estimated from the position and velocity estimates using \ref{eqn:param}.

We adopt TDoA-based localization as used in past work \cite{tiemann2016atlas}. Consider a tag positioned at $\textbf{x}_e$ and M anchor nodes with fixed locations $\textbf{X} = \{ \textbf{x}_1, \textbf{x}_2,\cdots, \textbf{x}_M \}$. We define the distances between the tag and the anchor nodes as $\{ D_1, D_2,\cdots, D_M \}$ and the range difference measurement at anchor node $i$ with respect to anchor node 1 by $\Delta_{i,1} = D_i - D_1$.
Then, the position of the a mobile tag $\textbf{x}_e$ can be determined using range and range difference measurements \cite{bensky2016wireless} by solving the equation given by (\ref{eqn:tdoa}):

\begin{equation}
    \textbf{x}_e = (\textbf{A}^T\textbf{A})^{-1}\textbf{A}^T (\textbf{b} + \textbf{c} D_1) 
\label{eqn:tdoa}
\end{equation}
where $\textbf{A} = \left[ \textbf{x}_1 - \textbf{x}_2, \cdots, \textbf{x}_1 - \textbf{x}_M \right]$, $\textbf{c} = \left[\Delta_{2,1}, \cdots, \Delta_{M,1} \right]^T$, $D_1 = \|\textbf{x}_e - \textbf{x}_1\|$, and 
\begin{equation}
    \textbf{b} =
    \begin{bmatrix}
    \|\textbf{x}_1\|^2 - \|\textbf{x}_2\|^2 + \Delta_{2,1}^2\\
    \vdots\\
    \|\textbf{x}_1\|^2 - \|\textbf{x}_M\|^2+ \Delta_{M,1}^2
    \end{bmatrix}.
\end{equation}

Assuming rectilinear motion for small fixed duration $T$, the position coordinates estimated from (\ref{eqn:tdoa}) are used to determine the average velocity. We apply batch linear regression to position coordinates and the corresponding time on a window size of $T$ seconds. Then, the computed slopes become the velocities of the nodes.  

Once the velocity is estimated, the initial position and velocity estimated for each pair can be used to estimate collision prediction parameters using (\ref{eqn:param}).

In this scheme, the system includes a receiving anchor node and synchronization node all with fixed known locations. 
\cite{bensky2016wireless}

\subsubsection{Friend based Collision Prediction}
In this approach, no anchor nodes or infrastructure exists.  Nodes can only measure range with respect to other mobile nodes called \textit{friend nodes}. 

One intuitive approach is to determine relative position using classical multidimensional scaling (MDS) which maps pairwise dissimilarities to high dimensional coordinates \cite{rajan2019relative}. However, using MDS for collision prediction is difficult for multiple reasons. First, it does not provide relative velocity info which is critical in collision prediction. Second, successive position estimates of a node can be translated, rotated, or flipped since there is no frame of reference. 

Past work has made some effort to estimate relative velocity within an MDS framework \cite{rajan2015joint}. 
However, this approach results in less accurate velocity estimates for noisy range measurements because it relies on the second order time difference of distance measurements.  Further, the relative velocity estimates  are, in general, from a different frame of reference from the MDS position estimates. 

\begin{figure*}[t]
     \centering
     \includegraphics[width=\columnwidth]{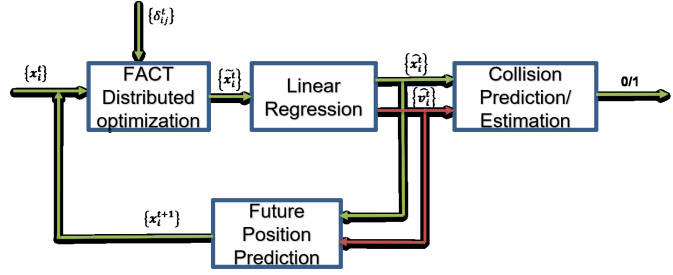}
     \caption{Collision detection based on FACT algorithm}
     \label{fig:fact_diagram}
 \end{figure*}

In this article, we propose the friend-based autonomous collision prediction and tracking (FACT) algorithm to estimate relative position and velocities in a distributed manner. 
Consider $N$ mobile nodes with their position at time $t$ given by $\textbf{X}^t = \{\textbf{x}_1^t, \textbf{x}_2^t, \cdots, \textbf{x}_N^t\}$. The pairwise distance measurements $\{\delta_{ij}^t\}_{i,j=1}^N$ between the nodes correspond to the actual Euclidean distances $\{d_{ij}^t\}_{i,j=1}^N$, i.e.
\begin{equation}
    d_{ij}^t = d(\textbf{x}_i^t, \textbf{x}_j^t)  =\| \textbf{x}_i^t -\textbf{x}_j^t \| = \sqrt{(\textbf{x}_i^t -\textbf{x}_j^t )^T(\textbf{x}_i^t -\textbf{x}_j^t )}.
\end{equation}
FACT estimates relative positions by minimizing a cost function given by:  
\begin{eqnarray}
    \text{S} &=&   
    \sum_{i=1}^{N}\sum_{j\neq i}\sum_{t=1}^{T} w_{ij}^t\left( \delta^{t}_{ij} - d_{ij}^t\right)^2 
    \nonumber \\
     && + \sum_{i=1}^{N} \sum_{t=1}^{T} 
     r_i \|\textbf{x}_i^{t-1} + \textbf{x}_i^{t+1} -2\textbf{x}_i^t\|^2 
    \label{eqn:cost}
\end{eqnarray}
The first term in the cost function ensures that the estimated positions $\{\textbf{x}_i^t\}_i$ are at the correct relative distances from each other at time $t$. 
 The second term is introduced to penalize changes in velocity of node $i$ by counting the squared magnitude of the double difference in position, $\|(\textbf{x}_i^t- \textbf{x}_i^{t-1})-(\textbf{x}_i^{t+1}-\textbf{x}_i^t)\|^2$, as error. The objective is to minimize (\ref{eqn:cost}) with respect to $\{\textbf{x}_i^t\}$. Assuming i.i.d.\ distance measurements, the cost function can be expressed
 as a sum of local cost functions:
 \begin{equation}
     \text{S} = \sum_{i=1}^N \text{S}_i + c
 \end{equation}
where $c$ is a constant, and with local cost function defined as:
\begin{eqnarray}
    \text{S}_i &=&  
    \sum^N_{\substack{{j=1}\\{j\neq i}}}
    \sum_{t=1}^{T} w_{ij}^t 
    \left( \delta^{t}_{ij} - \|\textbf{x}_i^t -\textbf{x}_j^t\|\right)^2
    \nonumber \\
     && + 
     \sum_{t=1}^{T} 
       r_i\|\textbf{x}_i^{t-1} + \textbf{x}_i^{t+1} -2\textbf{x}_i^t\|^2.
    \nonumber
\end{eqnarray}
We note that the local cost function $\text{S}_i$ depends only on measurements at node $i$ and positions of friend nodes. Minimizing the cost function can be done iteratively by using a quadratic majorizing function. We find a majorization function $T_i(\textbf{x}_i, \textbf{y}_i) \leq S_i(\textbf{x}_i)$ for all $\textbf{y}_i$ \cite{costa2006distributed}:
\begin{eqnarray}
    T_i(\textbf{x}_i, \textbf{y}_i) &=&
    \sum^N_{\substack{{j=1}\\{j\neq i}}}\sum_{t=1}^{T}
    {w_{ij}^t\left( (\delta^{t}_{ij})^2 + (d_{ij}^t)^2\right)} 
    \\
    && + \sum_{t=1}^{T}r_i\|\textbf{x}_i^{(t-1)} + \textbf{x}_i^{(t+1)} -2\textbf{x}_i^t\|^2 
    \nonumber \\
     &&  + \sum^N_{\substack{{j=1}\\{j\neq i}}}\sum_{t=1}^{T} w_{ij}^t \frac{ \delta^{t}_{ij}}{ d^{t}_{ij}} (\textbf{x}_i^t -\textbf{x}_j^t)^T(\textbf{y}_i^t -\textbf{y}_j^t). \nonumber
\end{eqnarray}
Since  $T_i(\textbf{x}_i, \textbf{y}_i)$ is quadratic function of $\textbf{x}_i$, the position can be determined by finding $\{\textbf{x}_i^t\}$ to minimize $T_i$ iteratively.  The distributed optimization is given below in Algorithm \ref{algo:fact}. The overall algorithm is summarized by in Figure \ref{fig:fact_diagram}.



\begin{algorithm}[H]
\begin{algorithmic}[1]
\State{\textbf{Inputs:} $\{\delta^{t}_{ij}\}$,$\{w_{ij}^t\}$, $\epsilon$, $\{r_i\}$, $\{x_i^t\}$,initial condition $\textbf{X}^{(0)}$}
\State {\textbf{Initialize:} $k = 0$, $\text{S}^{(0)}$, compute $a_i$ from equation (\ref{eqn:a_const})}
 \Repeat
      \State {$ k\gets k+1$}
      \For{$t = 1, \ldots, T$}
        \For{$i = 1, \ldots, N$}
            \State  {compute $\textbf{b}^{(k-1)}_i$ from  (\ref{eqn:b_const})} 
            \State {${\textbf{x}^t_i}^{(k)} = a_i \left(r_i {\textbf{x}^t_i}^{(k-1)} + \textbf{X}^{(k-1)}\textbf{b}^{(k-1)}_i\right)$}
            \State {$ \text{S}^{(k)} \gets \text{S}^{(k)} - \text{S}^{(k-1)}_i +\text{S}^{(k)}_i$}
            \State {communicate ${\textbf{x}^t_i}^{(k)}$ to friend nodes}
            \State {communicate ${\text{S}}^{(k)}$ to node $ (i+1)\mod{N}$}
        \EndFor
    \EndFor
    \Until{$\text{S}^{(k-1)} - \text{S}^{(k)} < \epsilon$}
\end{algorithmic}
 \caption{FACT distributed optimization}
 \label{algo:fact}
\end{algorithm}

Where
\begin{equation}
    a_i^{-1} = \sum^N_{\substack{{j=1}\\{j\neq i}}}\xoverline{w_{ij}} +r_i 
    \label{eqn:a_const}
\end{equation}
and $\textbf{b}^{(k)}_i = [b_1, b_2, \cdots b_N]^T$ is a vector given by
\begin{equation}
    \begin{aligned}
        b_j &= \xoverline{w_{ij}}[1-\delta_{ij}^t/d_{ij}^t(\textbf{X}^{(k)})], \qquad \mbox{for } j\neq i\\
        b_i &= \sum^N_{\substack{{j=1}\\{j\neq i}}}\xoverline{w_{ij}}\delta_{ij}^t/d_{ij}^t(\textbf{X}^{(k)}).
    \end{aligned}
    \label{eqn:b_const}
\end{equation}

Successive estimated coordinates in the algorithm are then used to determine the relative velocity of each node by applying linear regression on them with respect to time. Note that the algorithm starts with initialized positions $\textbf{X}^{(0)}$ of nodes.  We use MDS to initialize these positions at the start of tracking. For following rounds, we use projected coordinates, using the prior time coordinate and velocity estimates, to initialize the algorithm. We set the the first node at the origin, i.e $\textbf{x}_{ij}^t = \mathbf{0}$ and the weight matrix is given as:
\begin{equation}
  w_{ij}^t =
    \begin{cases}
      1 & \text{if $i \neq j$}\\
      0 & \text{otherwise}
    \end{cases}       .
\end{equation}
Although it may be possible to improve the algorithm by adaptively setting $w_{ij}^t$, we leave such approaches to future work.
\section{Multi-node Ranging}

Global navigation satellite systems (GNSS) are the most widely used positioning systems for outdoor applications. These systems fail in indoor and urban settings due to high attenuation and multipath fading. In recent years, time-of-flight measurements from impulse radio UWB (IR-UWB) transceivers have demonstrated to have cm-level ranging accuracy, even in severe multipath environments.  IR-UWB systems use pulse widths on the order of nanoseconds and an RF bandwidth on the order of GHz in order to achieve precise timestamping, which enables accurate ranging.  Past work in IR-UWB ranging and positioning is mostly limited to either single pair ranging or ranging between individual tag and an infrastructure of anchor nodes. In this section, we introduce a system that can measure IR-UWB range measurements between multiple mobile nodes using the minimal number UWB message exchanges, which works with or without infrastructure nodes.

\subsection{Hardware}
We implement multi-node ranging using commercial IR-UWB transceivers.  Every node is equipped with an ARM Cortex-M4 processor connected to an IR-UWB transceiver, a narrowband transceiver, and a shared voltage controlled temperature compensated crystal oscillator (VCTCXO). We use a Decawave DW1000 IR-UWB radio which supports IEEE 802.15.4a, and provide the timestamp of the arrival of the first arriving path \cite{yavari2014ultra}.  For clock synchronization, we adopt radio frequency synchronization (RFS) method proposed in \cite{luong2018stitch} where one node periodically broadcasts a reference unmodulated signal using a TI CC1200 narrowband transceiver and the other nodes compute carrier frequency offset and adjust their VCTCXO frequency. The CC1200 transceiver is also used to exchange data between nodes. Frequency synchronization is performed less frequently compared to UWB message exchanges, and the narrowband transceiver is programmed to communicate the UWB message timestamps.

\subsection{Multi-node Ranging Protocol}

Typically, two-way ranging requires at least two message exchanges for every pair of nodes with frequency-synchronized clocks \cite{yavari2014ultra}. For $N$ nodes, this would require $N(N-1)$  message exchanges to compute ranges for all pairs. In this article, we present a novel multi-node ranging protocol which requires only $N$ messages per ranging cycle. 

\begin{figure}
    \centering
     \includegraphics[width=0.5\columnwidth]{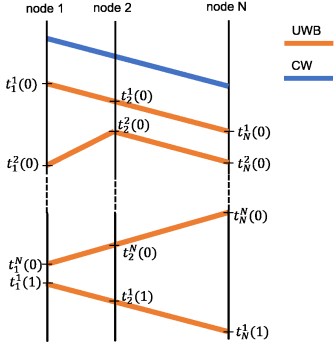}
    \caption{Multi node two-way ranging protocol}
    \label{fig:multinode_friend}
\end{figure}

\begin{figure}
    \centering
     \includegraphics[width=0.5\columnwidth]{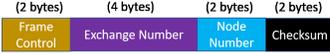}
    \caption{UWB packet format}
    \label{fig:uwbpkt}
\end{figure}

Figure \ref{fig:multinode_friend} illustrates our multi-node ranging protocol. At the start of the protocol, node 1 transmits a reference unmodulated signal over a narrowband channel for clock frequency synchronization. Up on receiving the narrowband signal, every other node applies RFS \cite{luong2018stitch} to synchronize its local VCTCXO. Next, starting from node 1, each node takes a turn transmitting a UWB packet, until the last node transmits its packet, completing the cycle. The UWB packet format, shown in Figure \ref{fig:uwbpkt}, includes an exchange number and the transmitter id. Each node stores  its most recent transmit and receive timestamps, and  broadcasts them over the secondary narrowband radio. Frequency synchronization is performed once each 100 cycles.

To compute two-way range $r_{ij}$ between nodes $i$ and $j$ (when $i>j$) in cycle  $n$ is given as:
\begin{equation}
    r_{ij}[n] = \frac{c}{2} \left\{\left(t_j^i[n] - t_j^j[n-1]\right)- \left(t_i^j[n] - t_i^j[n-1]\right)\right\} 
\end{equation}\label{E:r_ij_n}
where $t_j^j$ is the transmit timestamp at node $j$, $t_j^i$ is the receive timestamp at node $j$ for a message sent from node $i$, and $c$ is the speed of light. Note that $t_j$ and $t_j$ refer to timestamps recorded on different frequency-synchronized clocks, and thus may differ by a constant offset.  However, (\ref{E:r_ij_n}) uses only the differences in timestamps at $j$ and $i$, and thus any constant offset is removed.  Without frequency synchronization, (\ref{E:r_ij_n}) would be inaccurate because the offset would be changing with time.

\begin{figure}
    \centering
     \includegraphics[width=0.5\columnwidth]{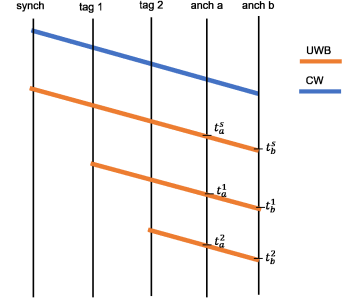}
    \caption{Anchor-based one-way ranging protocol to measure range differences}
    \label{fig:multinode_anch}
\end{figure}

We can use the same multinode ranging protocol to implement time-difference-of-arrival (TDOA) localization, which we do in this article for comparison.  In this case, we assign some static nodes to be the \textit{anchor} nodes, which are set to receive-only, i.e., recording receive time stamps but never transmitting.  In TDOA, full time synchronization between anchor nodes is necessary \cite{yavari2014ultra}.
We adopt the TDOA synchronization method of \cite{tiemann2016atlas} which adds a node with known location called a \textit{synch node} which broadcasts a UWB packet at the start of every cycle as shown in Figure \ref{fig:multinode_anch}. Unlike past work, we additionally use achieve frequency synchronization without wires by configuring the synch node to transmit an unmodulated narrowband signal after every 100 cycles, to which the anchors synchronize their TCVCXO.

For anchors $a$ and $b$, and a synch node with coordinates $\textbf{x}_a$, $\textbf{x}_b$  and $\textbf{x}_s$, respectively, the range difference $\Delta r_{ab}^i$ between anchor $a$ and anchor $b$ with respect to a given tag $i$ at cycle $n$ is given by:
\begin{eqnarray}
     \Delta r_{ab}^i[n] &=& c \left\{ (t_a^i[n] - t_a^s[n]) -(t_b^i[n] - t_b^s[n])\right. + \nonumber \\  && 
     \left. \|\textbf{x}_a -\textbf{x}_s\| - \|\textbf{x}_b -\textbf{x}_s\| \right\}.
     \label{eqn:r_ab}
\end{eqnarray}
The range difference measurements are then used to locate a tag based on TDOA-based localization as given by (\ref{eqn:tdoa}).
\section{Experiments}

To validate FACT and to compare it to other collision prediction. methods, we conduct experiments involving floor robots in a research laboratory.  

\begin{figure}[t]
  \begin{center}
  	\includegraphics[width=0.7\columnwidth]{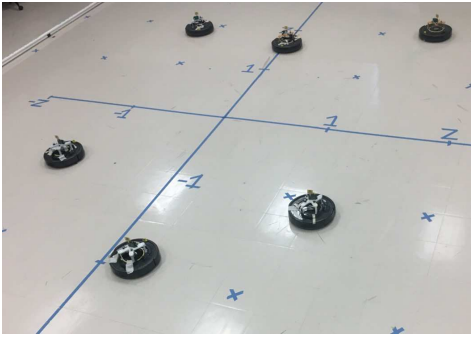}
  	\caption{Measurement experiment environment with six mobile nodes moving within our 7 m by 7 m area, with blue tape used to define coordinate axis in meters.  One PVC pipe ``wall'' is visible at top left.}
  	\label{fig:setup}
  \end{center}
\end{figure}

We show the experimental environment in Figure \ref{fig:setup},  a total of six iRobot Create robots are deployed in 7 m by 7 m empty area in a research laboratory. Our hardware is placed on top of the robot, with IR-UWB antenna in the center of the iRobot. The area surrounding the robots has significant RF clutter, including desks, desktop computers, soldering stations, and RF and digital measurement equipment.  The empty part of the lab where the experiments are conducted are lined with a set of OptiTrack optical tracking cameras.  PVC pipe is used at the edges as a ``wall'' to prevent the robots from leaving the 7 m square area.

The hardware we use on the iRobots include a UWB node for performing the ranging measurements, and a Raspberry Pi 3 processor that controls the robot motion. The robots are also tagged with four reflective markers so that the optical tracking system can track its position and orientation in the space.  Our code controls a robot to move, in a random direction, at a constant velocity of 0.5 m/s.  The robot continues until it hits an obstruction, either robot or wall, at which point it changes  direction randomly and starts a new constant velocity path. 

The OptiTrack motion capture system records ground truth data. It provides 3D node coordinates with mm-level accuracy using a set of 16 infra-red cameras at a 60 Hz rate.  The experiment involves two setups: 
\begin{enumerate}
    \item Setup 1: Six mobile nodes are programmed to measure two-way UWB ranges between each other at rate of 18 ranges per second for use in distributed collision prediction.
    \item Setup 2: The second setup involves five mobile nodes, three anchor nodes, and one synch node each positioned 30~cm above the floor at each corner of the experiment area. We measure one way range differences at each anchor node at a rate of 18 samples per second. 
\end{enumerate}
 The experiment is run for 18 minutes under each setup.

While our proposed method uses a distributed algorithm for collision prediction, in order to compare the results with other centralized algorithms, we capture ranging data at a receiver node directly connected via USB to a Dell XPS tower which performs the offline  processing. 

\section{Experimental Results}
We experimentally evaluate the performance of the proposed system both in ranging accuracy and in collision prediction performance. Ground truth coordinates obtained from OptiTrack system are used to generate collision labels and validate the results.
 
\subsection{Ranging Accuracy}
\begin{figure}
    \centering
     \includegraphics[width=0.7\columnwidth]{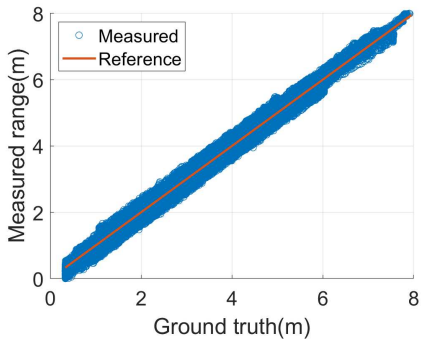}
    \caption{Ranging accuracy}
    \label{fig:gt_rng}
\end{figure}

The performance of range-based collision prediction relies on accurate  range estimates. First, we show the ranging accuracy of the system proposed in this paper. The system provides one-way range difference (\ref{eqn:r_ab}) and two-way range (\ref{E:r_ij_n}) measurements. In Figure \ref{fig:gt_rng}, we compare range estimates with with actual distances between nodes. Our system provides accurate range estimates with standard deviation of only 0.08~m. For one way range difference measurements, the system has a standard deviation 0.17~m.


\begin{figure}
    \centering
    \begin{minipage}{0.4\textwidth}
     \includegraphics[width=\columnwidth]{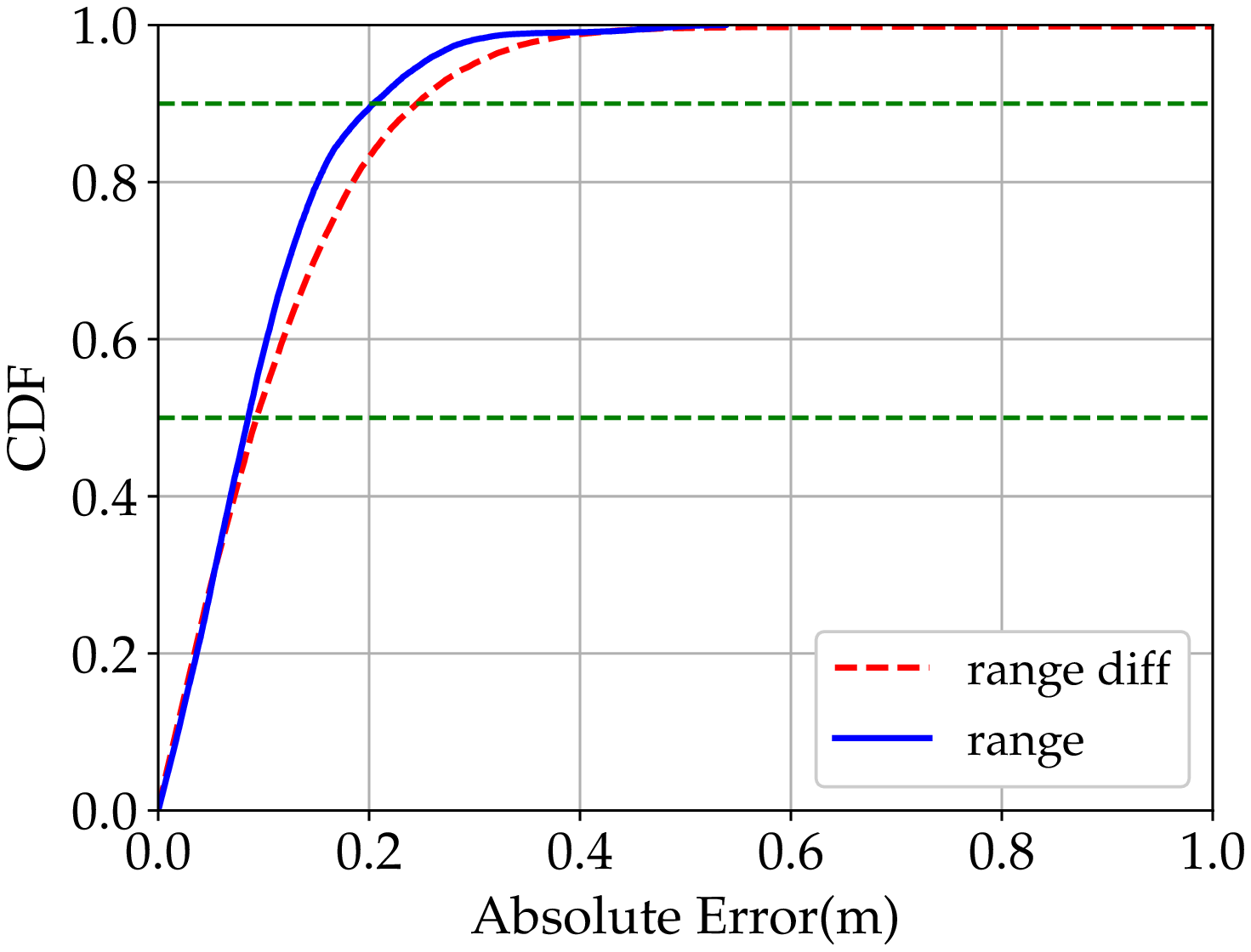}
     \end{minipage}
     \begin{minipage}{0.4\textwidth}
     \includegraphics[width=\columnwidth]{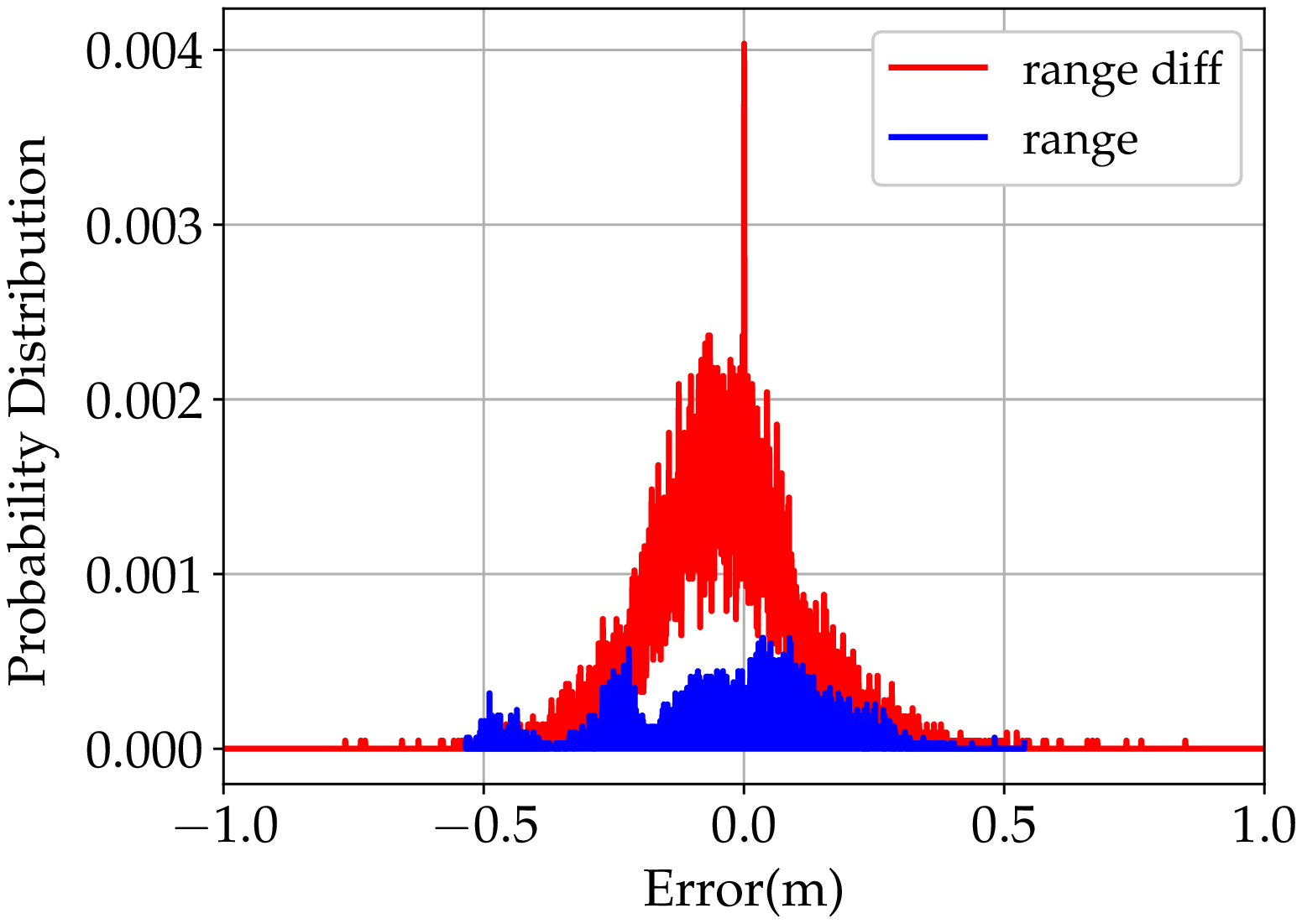}
     \end{minipage}
    \caption{Ranging error distribution}
    \label{fig:rng_cdf}
\end{figure}

Figure \ref{fig:rng_cdf} shows the cumulative distribution function of absolute range error for range and range difference measurements using our system. We note that higher accuracy is obtained with range estimates compared to one-way range differences. The root mean squared error (RMSE) for range estimation is 0.13~m whereas the RMSE for range differences is 0.21~m. The overall ranging accuracy of the system is summarized in Table \ref{tbl:rng_accuracy}.

\begin{table}[h]
\caption{Ranging accuracy for two-way ranges and one-way
range differences.}
\label{tbl:rng_accuracy}
\begin{tabular}{|l|c|c|c|c|}
\hline
                                                                                 & \multicolumn{2}{c|}{\cellcolor[HTML]{FFFFFF}{\color[HTML]{000000} \textit{\textbf{Error}}}} & \multicolumn{2}{c|}{\cellcolor[HTML]{FFFFFF}{\color[HTML]{000000} \textit{\textbf{Abs. Error}}}} \\ \hline
\rowcolor[HTML]{FFFFFF} 
\multicolumn{1}{|c|}{\cellcolor[HTML]{FFFFFF}{\color[HTML]{000000} Measurement}} & {\color[HTML]{000000} Bias(m)}               & {\color[HTML]{000000} RMSE(m)}               & {\color[HTML]{000000} Median(m)}             & {\color[HTML]{000000} 90$^\text{th}$ Perc. (m)}             \\ \hline
2-way range                                                                      & 0.022                                        & 0.13                                         & 0.085                                        & 0.2                                               \\ \hline
1-way range diff                                                                 & -0.045                                       & 0.21                                         & 0.093                                        & 0.24                                              \\ \hline
\end{tabular}
\end{table}














\begin{figure*}[tbhp]
\centering
\begin{subfigure}{0.45\textwidth}
     \centering
     \includegraphics[width=0.7\linewidth]{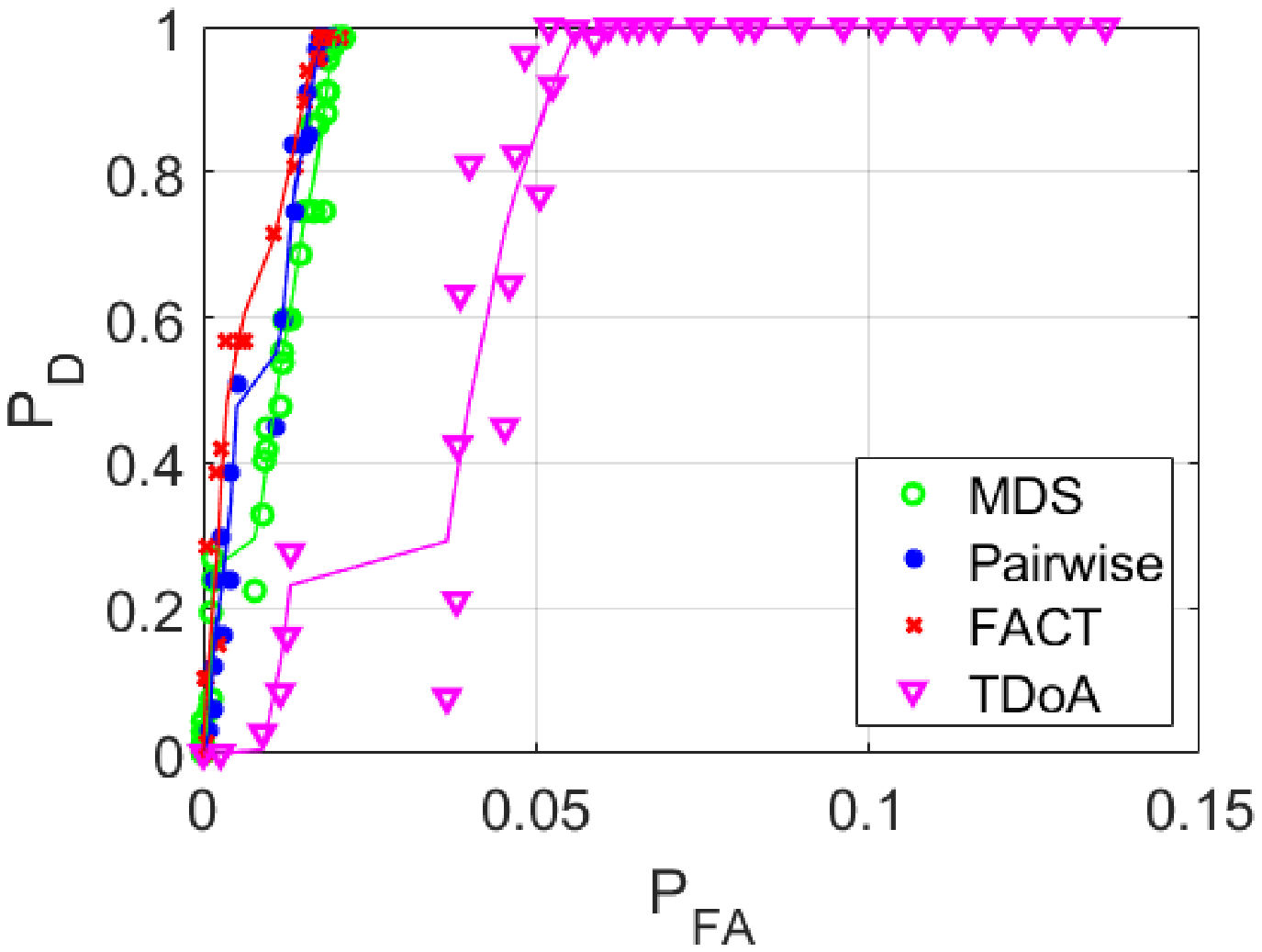}
     \caption{}
     \label{fig:combined_roc}
 \end{subfigure} %
{\Large$\xrightarrow{zoom}$}%
\begin{subfigure}{0.45\textwidth}  
    \centering
     \includegraphics[width=0.7\linewidth]{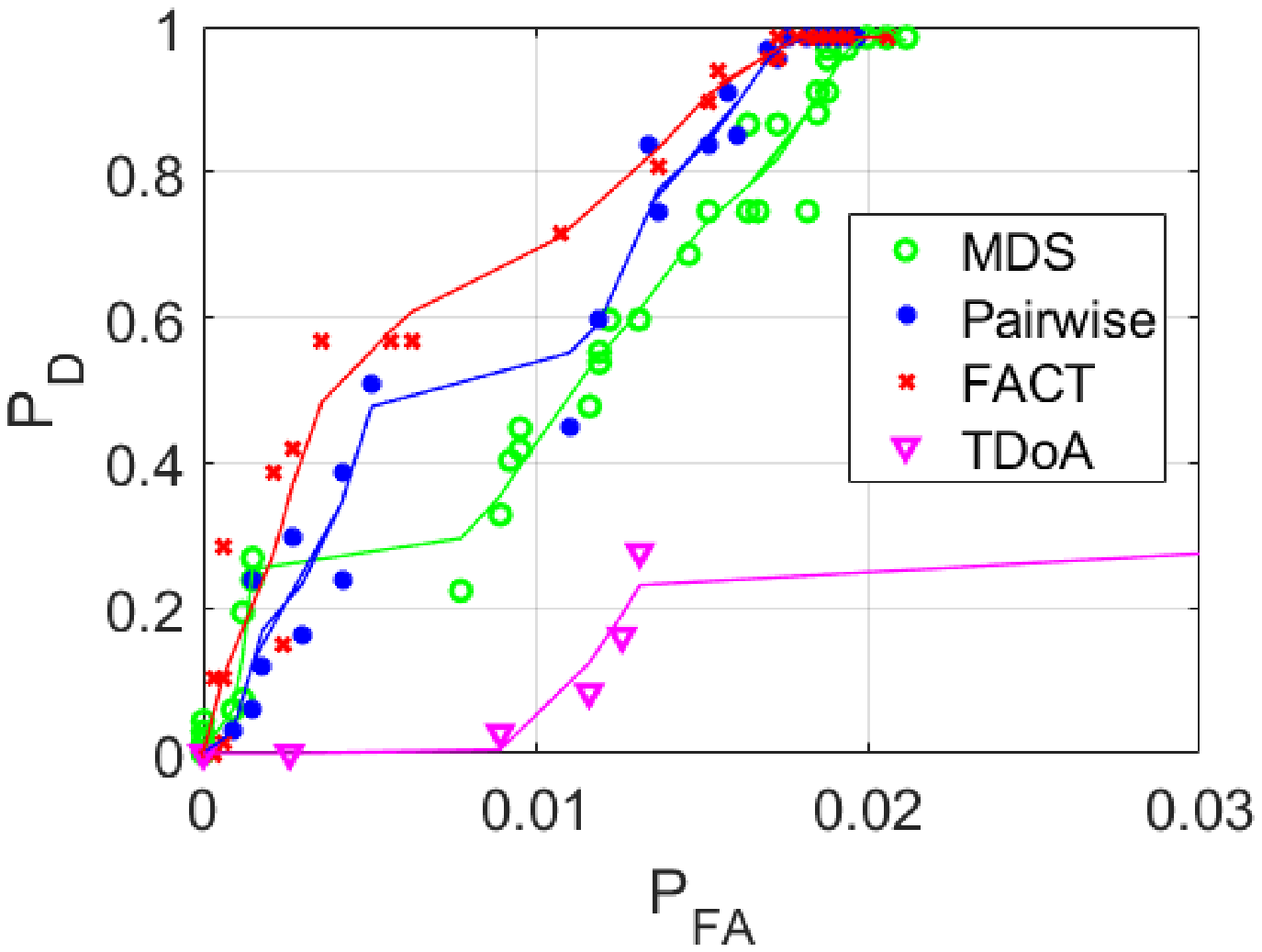}
    \caption{}
    \label{fig:combined_roc}
\end{subfigure} %
\caption{Receiver operating characteristic for collision prediction on experiment data where the solid lines are Lowess-smoothed representations. FACT achieves the highest probability of detection while TDoA method provides the lowest probability of detection at a given probability of false alarm.}
\label{fig:rng_roc}
\end{figure*}

\subsection{Collision Prediction} 
 To evaluate collision prediction performance of each proposed algorithm, we identify collision times using optical tracking data and and validate them using video captured during the experiment. We identify a total of 70 collisions from anchor-free and 60 collisions from anchor-based measurements. Assuming a player with radius $r$ and reaction time $\tau$ to detected collision, the system is determines oncoming collisions based on estimated values of the CP parameters (i.e, $d_m$, $t_m$, and $v$). 
 
 \begin{equation}
     \begin{aligned}
         \hat{d}_m &\lessgtr T_t = 2r + \epsilon_t,\\ 
         \hat{t}_m &\lessgtr T_d = \tau +  \sqrt{\frac{4r^2-\hat{d}_m^2}{\hat{v}^2}}+ \epsilon_d,
     \end{aligned}
 \end{equation}
where $\epsilon_d$ (m) and $\epsilon_t$ (s) represent real-valued constants which parameterize a ``buffer'' in range and time for the collision detector, allowing it to be more or less conservative. Then, the probability of false alarm $P_{FA}(T_t, T_d)$ is controlled with the threshold values $\{T_t, T_d\}$. For our analysis, we use a reaction time $\tau=1$ s and the diameter of the each robot $2r = 0.34~m$.


We evaluate the methods proposed in this paper with experiment data. Figure  \ref{fig:rng_roc} shows the receiver operating characteristic (ROC) curve in which the probability of detecting oncoming collision is given as the function the probability of false alarm (PFA).  We compare four different methods: our position and velocity adaptation of the multi-dimensional scaling (MDS) method, the pairwise detector, the time difference of arrival (TDOA) positioning-based detector, and the proposed FACT method.  We note that distributed methods, including MDS \cite{rajan2019relative}, pairwise and FACT  methods, generally provide better collision prediction performance followed by the centralized TDOA method.  

It is shown that collision prediction based on TDOA provides the worst ROC curve with lowest probability of detection (PD) for any probability of false alarm. It achieved a PD of 90\% at a higher PFA of 5\% compared to the same PD achieved by the other methods only with less  than 2\% PFA. We note that the accuracy of TDOA based position estimates decreases when agents are positioned near the boundary of the area defined by the convex hull of the anchors.  In addition, as shown in Table \ref{tbl:rng_accuracy}, one-way range difference measurements used in TDOA approach has higher errors compared to two-way range measurements used in distributed collision prediction algorithms.

In our evaluation, distributed methods including pairwise, MDS-based and FACT methods achieved higher probability of collision of prediction at any given PFA. FACT outperforms all other methods, attaining a PD of 90\% at 1.5\% PFA. The MDS-based collision prediction depends on first finding relative position and relative velocity from squared ranges and their second order time derivatives. For noisy measurements, this sometimes leads to imprecise position and velocity estimates and hence lower PD. The pairwise regression-based method \cite{abrar2020demo}, although achieves higher probability of collision prediction, its performance declines severely for with highly noisy measurements.  

\begin{figure*}[tbhp]
\centering
\begin{subfigure}{0.45\textwidth}
     \centering
     \includegraphics[width=0.7\linewidth]{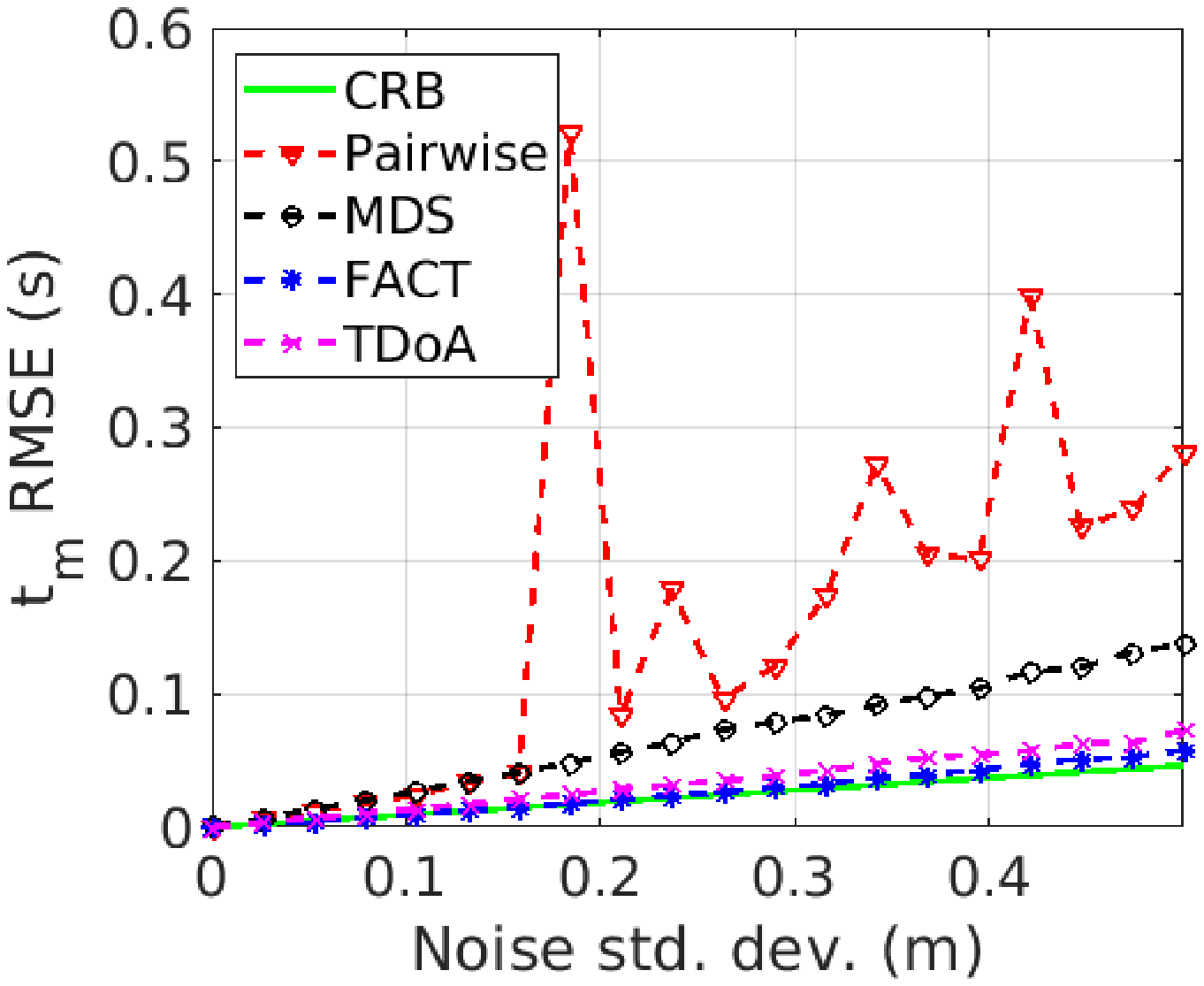}
     \caption{RMSE for $t_m$}
     \label{fig:tm_sim}
 \end{subfigure} %
 \quad
\begin{subfigure}{0.45\textwidth}  
    \centering
     \includegraphics[width=0.7\linewidth]{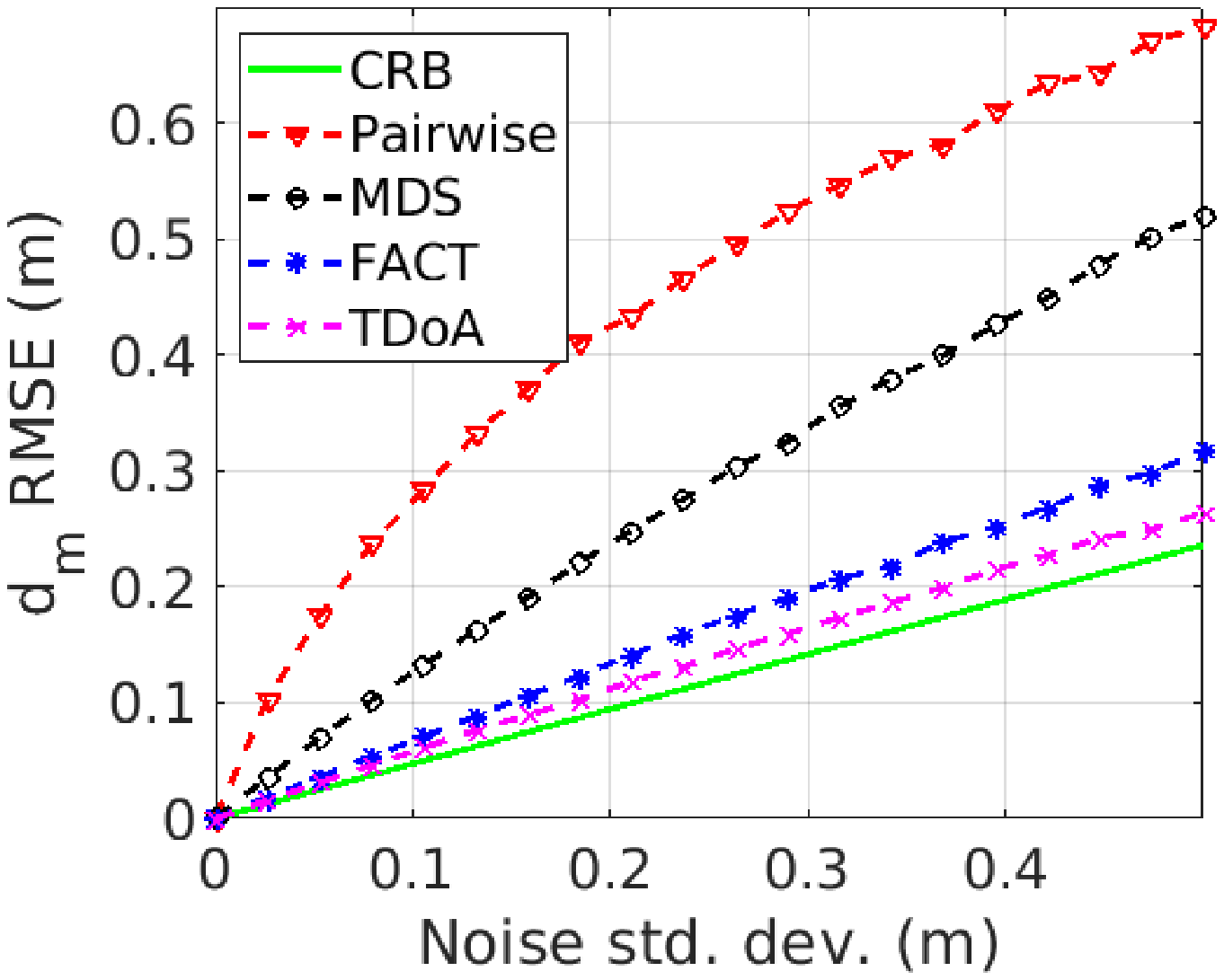}
    \caption{RMSE for $d_m$}
    \label{fig:dm_sim}
\end{subfigure} %
\caption{RMSE for CP parameters as a function of ranging noise based on simulation 150 different geometry and 100 iterations.}
\label{fig:dm_tm_rmse}
\end{figure*}

FACT leverages measurements from multiple pairs to enhance PD at lowest PFA. FACT has the potential to achieve even better performance with extended test scenario with more number of mobile agents.
Since the maximum speed of the robots used in the experiments is limited to 0.5 m/s, we also evaluate the methods using simulated data at a maximum speed of 10 m/s. The simulation involves 150 different geometries in 100 trials  with 4 anchors and 6 sensor nodes positioned in 2D as used in Section \ref{sec:est_bound}. Figure \ref{fig:dm_tm_rmse} compares the RMSE of estimates of the CP parameters, i.e $t_m$ and $d_m$, as a function of ranging noise for the methods proposed in this paper. The results show that FACT provides the lowest error for tm whereas TDOA based approach provides the lowest error for $d_m$. It is shown that the estimation error for the pairwise regression-based approach is considerably higher than all other methods. FACT maintains lower error for the CP parameters that is comparable to the CRB, further proving its potential in accurate collision prediction with noisy measurements while avoiding the need for infrastructure. 

We believe that FACT can be further enhanced with more sensor data including inertial measurements to extend linear motion assumption. However, this is beyond the scope of this article.

\section{Related Work}
\label{sec:related}

Several sensors and algorithms have been proposed  to predict collisions between autonomous vehicles, for example, swarms of drones or autonomous cars.  Various sensors such as lidar, radar, and visual sensors, are used to obtain raw information about events in the surrounding environment, which is followed by prediction of impending collisions.

{\vspace{0.1in} \noindent \bf Collision Detection Sensors:} include visual and acoustic sensors \cite{chellappa2004vehicle}, Doppler radar \cite{viquerat2008reactive}, tactile sensors \cite{ji2016flexible}, and Ultra Wideband (UWB) radar \cite{gresham2004ultra}. While some systems allow periodic exchange of position and velocity information between agents \cite{boivin2008uav, parker2007cooperative, xiang2014research}, other collision prediction systems are non-cooperative.  Active sensors including radar and lidar systems generally rely on continuous or pulsed electromagnetic wave transmissions to measure the ranging and Doppler information. Modern cars include a radar-based collision warning system called adaptive cruise control (ACC) \cite{marsden2001towards}.  Passive optical sensors or cameras are usually affected by changes in illumination and complex outdoor settings \cite{van2005vehicle, sun2004road}. With the exception of depth cameras, vision-based systems do not provide direct range information. Compared to optical/vision based collision prediction sensors, radar systems such as frequency-modulated continuous wave (FMCW) radars perform well under various light and weather conditions.

{\vspace{0.1in} \noindent \bf Collision Detection Algorithms:} use sensor data to calculate trajectory and determine distances between cooperating moving agents. 
 Vision-based systems typically employ a video feed from one or more cameras to track the trajectory of a moving agent using computer vision techniques \cite{sun2004road}. However, tracking many objects using video analysis can be computationally inefficient \cite{lin1997collision}. 
In robotics, moving agents can be tasked with path planning in robot swarms, where collisions can be avoided by employing motion planning in which each agent determines a preferred path that avoids collisions \cite{hennes2012multi}. For autonomous agents that are not part of the same ``swarm'', path planning may not be sufficient and collisions can be predicted by estimating one’s trajectory from relative kinematics such as relative position and relative velocity with respect to other nearby autonomous agents. 

Several positioning methods have been proposed in the literature. Depending on the type of sensors used, relative position and relative velocity are determined.  In radar systems, range and Doppler information is used to estimate relative position and relative velocity between moving agents.  Compared to most radar based measurements, UWB based ranging has proven to provide cost effective yet accurate ranging at higher update rate \cite{zafari2015microlocation}. 

Range based localization has been extensively studied. Common anchor-based localization methods such as those based on Time Difference of Arrival (TDoA) and Time of Arrival (TOA) measurements require careful infrastructure setup. When there are no nodes with known locations, anchor-free solutions like multi-dimensional scaling (MDS) solutions are used to map pairwise distance measurements into a geometry of nodes which generated them \cite{zhao2020graphips, wang2018formation, beck2014anchor}. However, tracking nodes with MDS to predict collision is difficult as MDS solutions are only up to rotation, reflection and translation. In \cite{rajan2019relative}, joint relative position and relative velocity estimation method is proposed in which relative velocity is determined from second-order time derivative of simulated range measurements and its accuracy is highly sensitive to ranging noise. Moreover, this method is centralized and unscalable which limits its use for practical applications.  In addition, collision prediction using UWB ranging has not completely addressed in prior work. This article provides a distributed solution for estimation of relative position and relative velocity from ranging measurements and applies them in collision prediction.


\section{Conclusion}
\label{sec:conc}

This article presents a novel approach to predict collision between moving agents, such as autonomous vehicles, that can measure and share distance measurements with each other and use them to predict that the two vehicles will collide. We develop a ranging system for measuring the range between every pair of $N$ nodes based on UWB transceiver measurements which requires only $N$ transmissions. We propose and evaluate a novel distributed collision prediction method that does not require infrastructure. We test the performance of the proposed method in theoretical analysis and experiments involving autonomous floor robots.  We compare performance to that of three alternate collision detection methods.  We show that our FACT algorithm outperforms infrastructure based methods and two alternate infrastructure-free methods. The FACT method allows collision detection between vehicles with low-cost hardware, without infrastructure, while simultaneously providing reliable collision prediction among mobile agents.

\section*{Acknowledgment}
This material is based upon work supported by the US National Science Foundation under Grant No. \#1622741.

\balance
\bibliographystyle{IEEEtran}

\end{document}